\documentclass[sigconf]{acmart}

\usepackage{xspace}
\usepackage{enumitem}
\usepackage{footmisc}
\usepackage{subcaption}
\usepackage{graphicx}
\usepackage{amsmath}
\usepackage{colortbl} 
\usepackage{soul}
\usepackage{balance}

\copyrightyear{2025}
\acmYear{2025}
\setcopyright{cc}
\setcctype{by-sa}
\acmConference[IMC '25]{Proceedings of the 2025 ACM Internet Measurement
Conference}{October 28--31, 2025}{Madison, WI, USA}
\acmBooktitle{Proceedings of the 2025 ACM Internet Measurement Conference
(IMC '25), October 28--31, 2025, Madison, WI, USA}
\acmDOI{10.1145/3730567.3764484}
\acmISBN{979-8-4007-1860-1/2025/10}

\ccsdesc[500]{Networks~Network measurement}
 
\begin{document}
  
\title[LACeS: An Open, Fast, Responsible, and Efficient Longitudinal Anycast Census System]{%
  LACeS: An Open, Fast, Responsible, and Efficient\\Longitudinal Anycast Census System
}
 
\newcommand{\ie}{\textit{i.e.}}
\newcommand{\etc}{\textit{etc.}}
\newcommand{\eg}{\textit{e.g.}}
 
\author{Remi Hendriks}
\affiliation{
  \institution{University of Twente}
  \city{Enschede}
  \country{The Netherlands}
}
\email{remi.hendriks@utwente.nl}
\orcid{0009-0008-1578-668X}

\author{Matthew Luckie}
\affiliation{
  \department{CAIDA}
  \institution{UC San Diego}
  \city{La Jolla}
  \state{CA}
  \country{USA}
}
\orcid{0000-0002-3872-4624}

\author{Mattijs Jonker}
\affiliation{
  \institution{University of Twente}
  \city{Enschede}
  \country{The Netherlands}
}
\orcid{0000-0001-5174-9140}

\author{Raffaele Sommese}
\affiliation{
  \institution{University of Twente}
  \city{Enschede}
  \country{The Netherlands}
}
\orcid{0000-0003-3484-9259}

\author{Roland van Rijswijk-Deij}
\affiliation{
  \institution{University of Twente}
  \city{Enschede}
  \country{The Netherlands}
}
\orcid{0000-0002-0249-8776}

\renewcommand{\shortauthors}{Remi Hendriks, Matthew Luckie, Mattijs Jonker, Raffaele Sommese, \& Roland van Rijswijk-Deij}

\def \manycasttwo  {MAnycast\textsuperscript{2}\xspace}
\def \manycast  {LACeS\xspace}
\def \CLI {\texttt{CLI}\xspace}
\def \Server {\texttt{Orchestrator}\xspace}
\def \Client {\texttt{Worker}\xspace}
\def \Clients {\texttt{Workers}\xspace}
\def \gcdark {GCD\textsubscript{\textit{LS}}\xspace}
\def \gcdfull {GCD\textsubscript{\textit{IPv4}}\xspace}
\def \AC {AC\xspace}
\def \ACs {ACs\xspace}

\begin{abstract}
IP anycast
replicates an address at multiple locations to
reduce latency and enhance resilience.
Due to anycast's crucial role in the modern Internet, earlier research introduced tools to perform anycast censuses.
The first, iGreedy, uses latency measurements from geographically dispersed locations to map anycast deployments.
The second, \manycasttwo,
uses anycast to perform a census of other anycast networks.
MAnycast\textsuperscript{2}'s advantage is speed and coverage
but suffers from problems with accuracy,
while iGreedy is highly accurate
but slower using author-defined probing rates and costlier.

In this paper we address the shortcomings of both systems and present \manycast (Longitudinal Anycast Census System).
Taking \manycasttwo as a basis, we completely redesign its measurement pipeline,
and add support for distributed probing, additional protocols (DNS over UDP, TCP SYN/ACK, and IPv6) 
and latency measurements similar to iGreedy.
We validate \manycast on an anycast testbed with 32 globally distributed nodes, 
compare against an external anycast production deployment, extensive latency measurements with RIPE Atlas
and cross-check over 60\% of detected anycast using operator ground truth that shows \manycast achieves high accuracy.
Finally, we provide a longitudinal analysis of anycast, covering 17+months, showing \manycast achieves high precision.
We make continual daily \manycast censuses available to the community and release the source code of the tool under a permissive open source license.
\end{abstract}

\keywords{Internet Measurement, Anycast, Internet Topology, Routing, IP}

\maketitle

\vspace{-0.005em}


\section{Introduction}

Anycast is a technique where an IP address is made available in
multiple, independent sites such that packets are routed to one of
several available sites~\cite{rfc4786}.
This technique is widely used to satisfy client requests from multiple
sites to ensure low-latency responses.
Anycast also allows for load-balancing by distributing traffic over
multiple sites.
Finally, it can enhance resilience by ensuring the availability of a
service in multiple sites that can absorb traffic in case one site
goes down.
Prominent examples of services that use anycast include the Domain
Name System (DNS)~\cite{dns-anycast} and Content Delivery Networks
(CDNs)~\cite{calder-imc-2015,koch-sigcomm-2021}.

Given its widespread use, determining how widely and for what services
anycast is used is vital to our understanding of the modern-day
Internet.
Several techniques have been developed to perform Internet-wide IP
anycast censuses.
The two most prominent are iGreedy (Cicalese et al., 2015~\cite{igreedy})
and \manycasttwo (Sommese et al., 2020~\cite{manycast}).
The first, iGreedy, uses latency measurements from geographically
dispersed Vantage Points (VPs) locations to infer anycast use.
iGreedy detects anycasted prefixes by looking for ``speed-of-light
violations'' -- latencies from multiple VPs to a target that are
physically impossible if the target is in a single location.
A census of the ping-responsive IPv4 address space using iGreedy takes
days~\cite{igreedy} on currently available community measurement
platforms, which limit packet rate, and critically relies on the VPs
having an accurate location.
The second technique, \manycasttwo~\cite{manycast}, involves probing
from multiple VPs using an anycast source address.
A target will send its responses to the VP that is closest in terms of
BGP routing.
A unicast target will therefore send responses from one site to a
single VP, whereas an anycast target's responses can be sent from
multiple sites to multiple VPs.
Because networks that can support anycast experiments are typically
well-provisioned, \manycasttwo can infer which IPv4 ping-responsive
prefixes are anycast in under 3 hours.
Unfortunately, \manycasttwo cannot infer {\em where} anycast sites
are, and can misclassify a prefix as anycast if routing from a
target changes during the measurement.

In preliminary work evaluating \manycasttwo, Sommese et al. suggested
a future pipeline where \manycasttwo identifies candidate anycast
prefixes for followup probing with iGreedy~\cite{manycast}.
In this work, we design and build a Longitudinal Anycast Census System
(\manycast) inspired by that model, that improves \manycasttwo's
accuracy through synchronized probing, and incorporates the
latency-based measurement algorithm from iGreedy.
\manycast's streamlined implementation outperforms \manycasttwo in
terms of speed, allowing for a census and accompanying latency-based
validation to be performed daily.
The contributions of this paper are that we:

\begin{itemize}[itemsep=1pt,topsep=1pt,labelindent=0.25em,leftmargin=*]
	\item Re-engineer \manycasttwo and iGreedy to improve accuracy, coverage (IPv6, UDP- and TCP-based probing) and more;
	\item Validate LACeS's performance against operator ground truth,
	measurements on an anycast production deployment, 
	commercial datasets,
	and latency measurements through RIPE Atlas;
	\item Release daily anycast censuses for the ping-responsive IPv4 address space and state-of-the-art IPv6 hitlists beginning Mar'24~\cite{gitcensus},
	accessible with a user-friendly dashboard~\cite{webcensus};
	\item Release the full source code for \manycast~\cite{gittooling}.
\end{itemize}

The remainder of this paper is organized as follows. 
First, we provide background on iGreedy and \manycasttwo
and discuss related work in \S\ref{background}.
In \S\ref{requirement} we discuss the requirements for \manycast.
Next, in \S\ref{methodology} we introduce the design and implementation of \manycast.
\S\ref{performance} discusses our extensive performance evaluation
followed with ground truth validation in \S\ref{groundtruth}.
Next, we provide a longitudinal analysis in \S\ref{sec:longitudinal}
and discuss \manycast and future steps in \S\ref{lessons}.
Finally, we conclude in \S\ref{conclusion}.

        
\begin{figure}[t]
    \includegraphics[width=0.71\columnwidth]{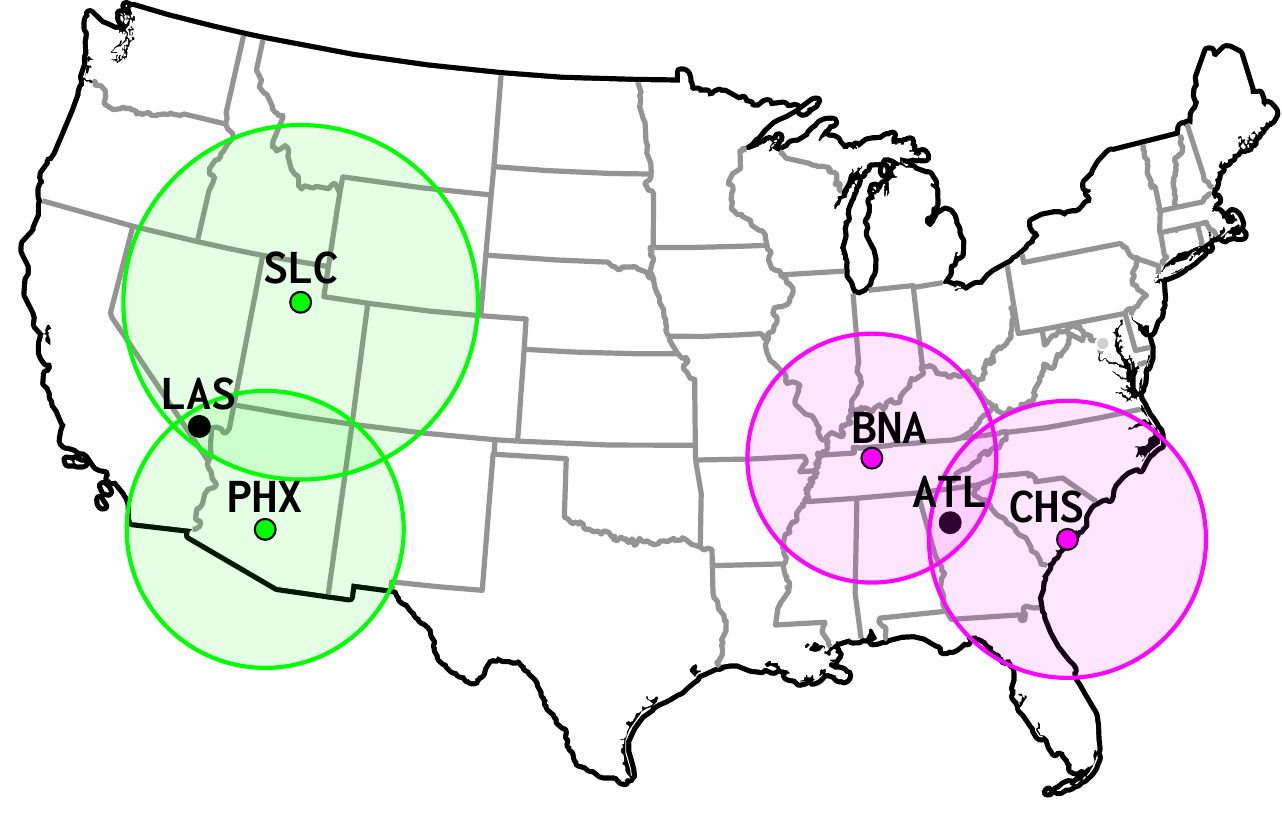}
  \caption{
    Using latency measurements from four VPs (SLC, PHX, BNA, CHS)
    for GCD-based anycast site (LAS, ATL) detection.  These latency
    measurements imply at least two sites.  iGreedy infers the location
    of each site using the highest populated city in the intersection.}
  \label{fig:igreedy}
\vspace{-1em}
\end{figure}

\section{Background and Related Work} \label{background}

This section provides background information on iGreedy 
and \manycasttwo.
We end with a discussion of other related work.

\subsection{iGreedy and the Great Circle Distance}
\label{igreedy}

iGreedy, introduced by Cicalese et al. in 2015~\cite{igreedy}, uses
the \textit{Great-Circle Distance} (GCD)
to determine the maximum geographical distance an Internet packet
travels based on its Round-Trip Time (RTT).
Figure~\ref{fig:igreedy} illustrates this principle -- by capturing
latency data from multiple VPs to a probed target, iGreedy determines
the areas in which a target could reside.
When there are non-overlapping circles (green and purple circles
in Figure~\ref{fig:igreedy}), iGreedy infers the target to be anycast.
iGreedy enumerates anycast sites by determining the minimum set of
areas 
and infers the geolocation by selecting the highest populated city in each
area.

The GCD approach relies on assumptions about the speed at which
packets travel.  iGreedy's default is the speed of light in fibre
($\sim$200,000\,km/s).
This estimate discards, \eg, delays due to buffering.
Therefore, the GCD approach can underestimate both the number of
anycasted IP prefixes and the number of sites per prefix.
In particular, it may fail to detect anycast deployed in close
proximity. 

\begin{figure}[t]
  \centering
  \includegraphics[width=0.71\columnwidth]{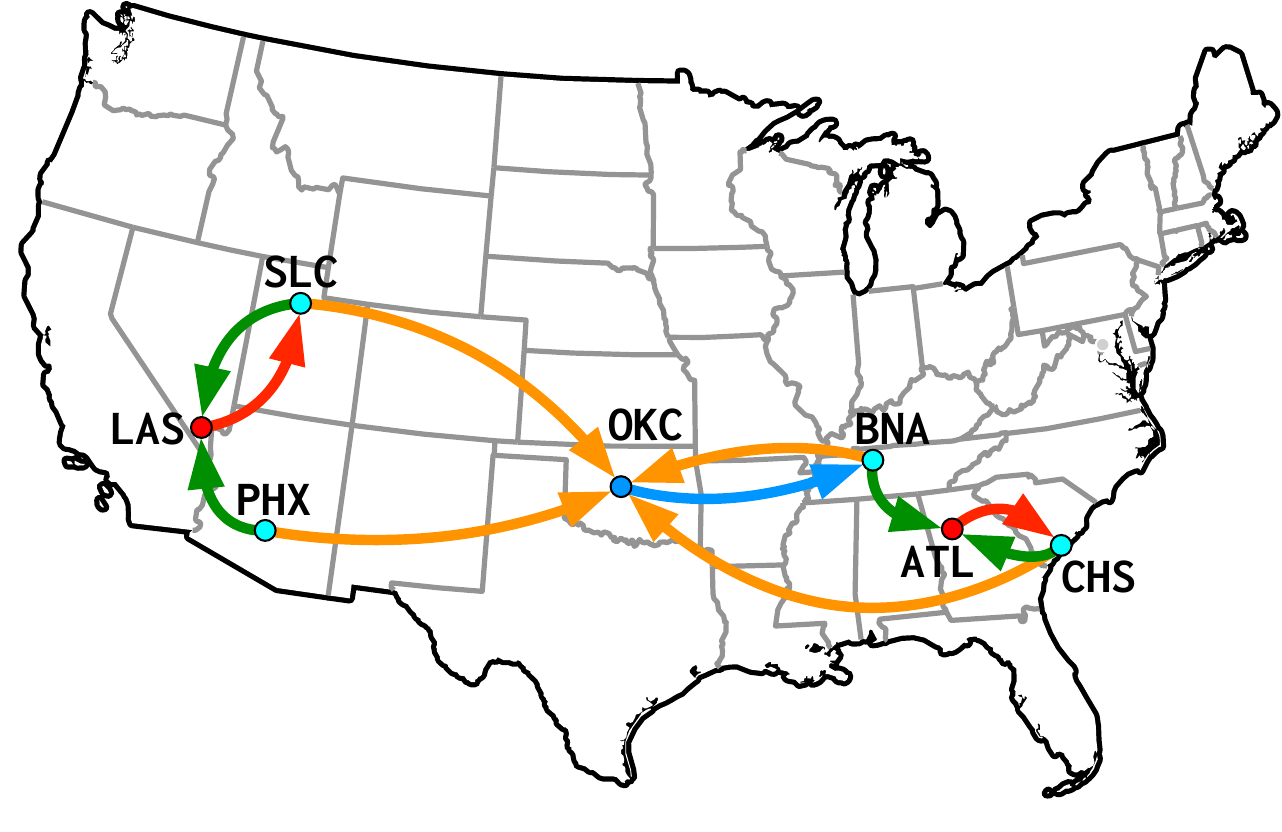}
  \caption{Using an anycast setup (light blue nodes -- SLC, PHX, BNA,
    CHS) to detect anycast. Probes to a unicast site (orange lines to
    OKC) results in responses to a single VP (BNA), while probes to
    anycast sites (green lines to LAS and ATL) result in responses to
    multiple VPs (red lines to SLC, CHS).}
  \label{fig:manycast}
\vspace{-1em}
\end{figure}

\subsection{\manycasttwo}
\label{manycast}

\manycasttwo~\cite{manycast} is based on an intuition by De Vries
et al.~\cite{verfploeter}.
De Vries et al.\ developed ``Verfploeter'' to map anycast
{\em catchments}, defined as the set of prefixes that send their traffic
to a given site of an anycast service.
They inferred catchments by pinging responsive addresses -- a {\em passive VP}
-- in each prefix using an anycast source address, noting the site
that responses arrive at.
When De Vries et al.\ tested catchment stability, they noticed that
some passive VPs were in the catchment of many sites.
They intuited that some of the prefixes associated with these passive
VPs were themselves anycast, and confirmed this with ground truth.

Figure~\ref{fig:manycast} illustrates how \manycasttwo leverages anycasted
VPs to detect anycast.
Probing an anycast address can lead to responses at multiple anycasted
VPs -- a \textbf{true positive (TP)}.
Probing a unicast address from multiple anycasted VPs can lead to responses
being received by a single VP -- a \textbf{true negative (TN)}.

However, there are cases where this intuition fails.
A \textbf{false positive (FP)} occurs when
there are equal-cost BGP paths to multiple anycasted VPs, so Equal-Cost
Multi-Path (ECMP) routers can send responses to multiple sites.
Furthermore, routing dynamics may cause a unicast target to respond to a
different anycast VP when a route changes between sequential
\manycasttwo probes.
A \textbf{false negative (FN)} occurs when a probed target's
anycast sites are in the catchment of a single anycasted VP, meaning
all sites of the target will respond to the same VP.
To minimize FNs the measuring anycast infrastructure must consist of
many topologically dispersed VPs, lowering the odds of a probed
anycast deployment being entirely in the catchment of a single VP.

Whilst \manycasttwo is able to detect anycast using few VPs, iGreedy
detects anycast with fewer FPs~\cite{manycast}, and is able to geolocate anycast sites
at a city-level granularity~\cite{igreedy}.
At the same time, iGreedy suffers FNs when sites are geographically close.
To find a middle ground between probing cost and reducing FPs, Sommese
et al. suggested a future pipeline where \manycasttwo identifies
candidate anycast prefixes for followup probing with iGreedy using a
much larger number of VPs~\cite{manycast}.
In this paper, we build this pipeline using both methodologies and
publicly share daily census results.



\subsection{Other related work}
\label{related_work}

\textbf{Anycast measurement methods} --
Anycast is used extensively for the DNS, particularly for the DNS root
servers.
Because it is useful to identify which site a client uses for
troubleshooting purposes, RFC~4892 specifies a mechanism for sites to
disclose their identity in a CHAOS-class DNS \texttt{TXT}
record~\cite{chaos_rfc}.
DNS server operators typically encode a location string in their
identity strings, so CHAOS TXT queries are useful when studying
anycast deployments.
In 2011, Fan et al.\ proposed ``ACE'' (Anycast Characterization and
Evaluation) that uses CHAOS records in combination with traceroute to
detect and enumerate anycast~\cite{chaos_traceroute}.
In 2013, Fan et al.\ improved the scope of this approach by
proposing Internet-class (IN) DNS records that can be requested using
existing recursive DNS infrastructure~\cite{chaos}.
They reported that 10\,k VPs were required for a recall of 80\% when
enumerating the number of anycasted replicas.
Because it relies on CHAOS queries, Fan et al.'s approach is DNS-specific.

In 2019, Bian et al.\ passively collected BGP announcements to detect
anycast by looking for geographically diverse upstreams
\cite{bgp_anycast}.
Their method is service-agnostic, as it does not rely on DNS records.
However, their methodology suffered from false positives due to remote
peering, because remote peering allows a unicast prefix to have
upstreams in geographically distant locations.

\textbf{Anycast censuses} --
In addition to developing iGreedy, Cicalese et al. performed monthly
iGreedy censuses of anycast between December~2015 and
May~2017~\cite{igreedy_longitudinal}.
In this work, we expand on this by providing a
fine-grained daily longitudinal dataset that will enable detailed
study of the IP anycast landscape.

Current censuses of anycast are
largely commercial.
\textit{BGPTools}, a popular platform amongst operators, offers a
public list of anycast prefixes using a methodology similar to
\manycasttwo~\cite{anycatch}
with four VPs as of Sep.\ '25.
Geolocation database providers like \textit{IPInfo}~\cite{ipinfo} have a latency-based
anycast classifications in their dataset.
Whereas these censuses only report anycast prefixes, we additionally enumerate
and geolocate anycast. 
We compare results with BGPTools and IPInfo in \S\ref{subsec:external}.



\section{System requirements} \label{requirement}
\manycasttwo suffers from a high false positive rate due to routing dynamics affecting anycast detection.
To overcome this issue, the authors proposed a two-step system.
However, they did not provide a single integrated solution to realize such measurements.
Furthermore, neither \manycasttwo nor iGreedy was designed for daily anycast censuses.
We wanted to build a system that:
i) combines anycast- and latency-based GCD measurements;
ii) is accurate and precise;
and iii) performs a responsible measurement in terms of probing budget.
We also want to release a continual daily census for the community
and release the toolchain
for others to build on.
To achieve these goals we set the following requirements:

\begin{enumerate}[label=R\arabic*,itemsep=1pt,topsep=1pt,labelindent=0.25em,leftmargin=*]
	\item \label{r:accuracy}\textbf{Accurate} - \manycasttwo suffers from false positives and false negatives. \manycast must minimise such cases, and
	must convey confidence in results through independently listing the classification for the anycast-based and GCD approach.

	\item \label{r:precision}\textbf{Precise} - Anycast detection errors between subsequent censuses must be minimized to support longitudinal measurements.
	In other words: \manycast must be precise over time.

	\item \label{r:responsibility}\textbf{Responsible} - \manycast must perform measurements with low impact on the platform it runs on and the Internet as a whole.
	\manycast must support low probing rates without impacting accuracy and precision.

	\item \label{r:coverage}\textbf{Multi-protocol} - iGreedy and \manycasttwo only support ICMPv4. \manycast must also support IPv6, transport layer protocols (TCP, UDP) and service-aware probing.
	
	\item \label{r:efficiency}\textbf{Performant} -
	\manycasttwo had poor scalability and iGreedy cannot perform a comprehensive IPv4 census in a single day.
	\manycast must complete both a hitlist-based IPv4 census and IPv6 census for multiple protocols in a single day.

	\item \label{r:reproducibility}\textbf{Reproducible} - Independent parties, that may have anycast deployments of various sizes and limited resource availability, must be able to run \manycast.
\end{enumerate}

\vspace*{-1mm}


\section{Design and Measurement Setup} 
\label{methodology}


We begin with a set of responsive addresses across the Internet
(\S\ref{hitlists}).
The \manycast probing system consists of two components: one
based on \manycasttwo (\S\ref{candidates})
that detects candidate anycast prefixes, and one based
on iGreedy that infers if those prefixes are anycast, and the location
of sites for anycast prefixes (\S\ref{pipeline}).

\subsection{Census Input}
\label{hitlists}
We use ISI's IPv4 hitlist~\cite{isi}, which ranks ping-responsive
addresses per /24 prefix, to select ping targets.
For DNS we merge the ISI hitlist with a set of authoritative name
server IPs obtained from the OpenINTEL project~\cite{openintel}.
We prefer name server IPs as representative addresses for a specific
/24 to maximize chances of probing an active DNS server.
Next, we created a TCP hitlist based on
Zmap~\cite{zmap} scans targeting the entire routable IPv4 address space.
These hitlists cover 6.0\,M /24s, with a union of 4.3\,M responsive (3.8\,M ICMP, 2.9\,M TCP, 0.2\,M DNS) across 73k ASes.

For IPv6 we use the public hitlist from TU Munich (TUM)~\cite{v6hitlist} combined with \texttt{AAAA} DNS records obtained from OpenINTEL.
We consider only the first /48 of large aliased prefixes
considered to be allocated to the same service,
to maintain reasonable probing burden and times.
This leaves 6.2\,M /48s, with a union of 5.6\,M responsive (5.3\,M ICMP, 4.3\,M TCP, 26k DNS) across 21k ASes.
Due to the longitudinal aspect of \manycast we update hitlists quarterly, in sync with ISI's hitlists~\cite{isi}, to maintain coverage.

We probe at a /24-IPv4-prefix and /48-IPv6-prefix granularity,
targeting a single address per prefix,
as these are likely to be the smallest prefix size to be propagated by BGP.
This reduces the probing burden and makes daily censuses feasible.
Whilst this assumption works well in most cases, Schomp et al. showed
that there may be rare cases of more specific
partitioning~\cite{anycast_partitioning}.
For example, hypergiants, such as Microsoft, announcing a /24 prefix
at their PoPs may assign part of a prefix to a replicated service,
whilst utilizing all other addresses in that prefix as unicast for
individual servers within their private backbone, we refer to this practice as \textit{partial anycast}.
While we acknowledge this limitation, we continue to probe at a /24
prefix granularity to keep our scanning rate low for responsible
scanning.
We address and quantify this limitation in \S\ref{sec:pfxsize}.

\subsection{Detecting Candidate Anycast Prefixes}
\label{candidates}


\begin{figure*}[tb]
  \centering
  \includegraphics[width=0.76\textwidth]{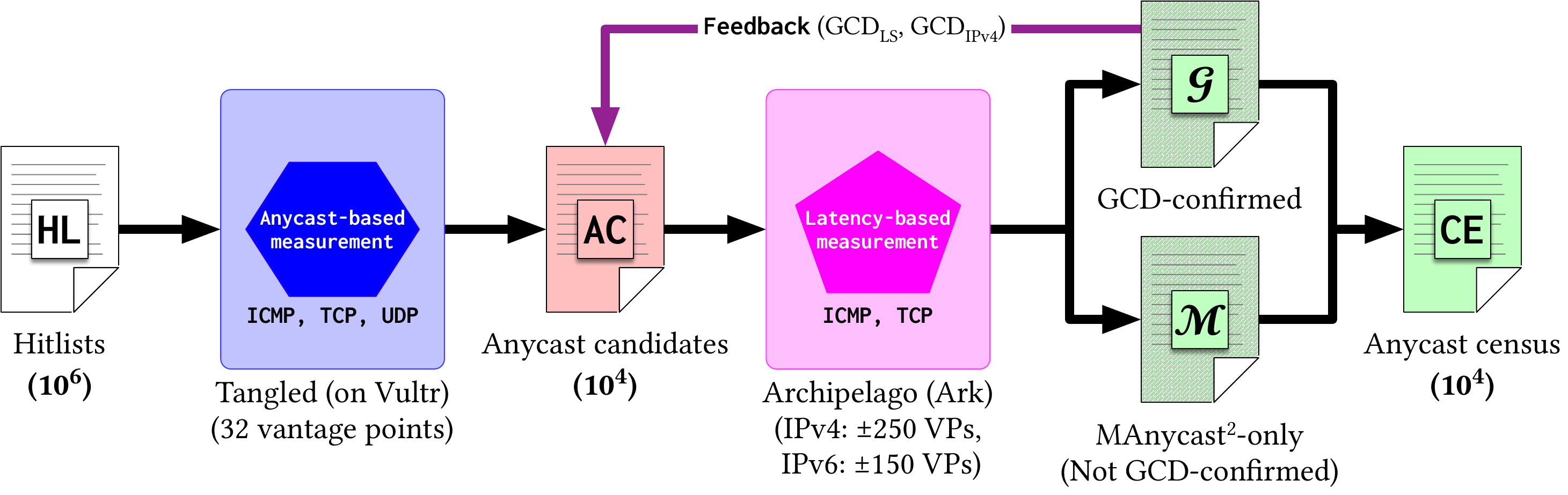}
  \caption{Overview of the measurement pipeline.}
  \label{fig:pipeline}
\end{figure*}

\subsubsection{Components}
Our anycast-based probing system consists of three components:

\vspace{0.25em}
\noindent
\textbf{\Server} -- Central controller for anycast probing, deployed
on-premise at our institution. 

\vspace{0.1em}
\noindent
\textbf{\CLI} -- Interface to create a measurement definition and instruct the \Server to perform a specific measurement.

\vspace{0.1em}
\noindent
\textbf{\Client} -- Deployed at the anycast sites to receive instructions from the \Server to probe the Internet.

We minimize the burden on \Clients by offloading all computation to
the \Server except for sending, receiving, and matching of packets.
The separate \CLI component allows measurements to be started from
multiple environments.
We deploy \Clients on TANGLED~\cite{tangled}, our anycast testbed that
is deployed on Vultr's cloud and uses of all its 32 sites, located in
19 countries on 6 continents~\cite{vultr_locations}.
We announce a /24 IPv4 prefix and a /48 IPv6 prefix from all sites.
In \S\ref{subsubsec:deployments} we motivate our decision to use Vultr.

\subsubsection{Daily measurement process}
When starting a measurement, the \CLI takes a measurement definition as input and forwards this to the \Server.
The \Server then instructs all \Clients that a measurement is starting,
including a definition of the measurement such that \Clients know what probes to send and listen for.
The \Clients then start listening.
Next, the \Server streams a list of IPs -- at the \CLI-defined probing rate -- to each \Client.
\Clients send out probes as they receive hitlist targets from the \Server.
Each hitlist target receives a single probe from each \Client, the \Server orchestrates it such that these probes are scheduled within a configurable offset after each other.
When performing an ICMP-based measurement at an offset of 1~second,
a target will receive a set of probes from the same source address one second after each other,
mimicking a regular ping sequence.
We encode the sending \Client ID and the transmission time in fields that are echoed in responses from targets.
For ICMP we use the ICMP payload,
for DNS we encode information in the domain name of the request,
and for TCP we use the acknowledgement number.

If a probed target is responsive,
it sends back a response for each received probe and these responses are routed to the nearest \Client,
except for the case of the probed target being anycast as explained in \S\ref{background}.
The \Clients capture responses to probes and ensure they belong to the ongoing measurement by checking the information encoded in the probe that is echoed by the target.
The \Client then 
creates a result consisting of information encoded in the original probe
and information such as the receive time and the receiving \Client ID.
Results are streamed back to the \Server that receives streams from all \Clients and aggregates results in a single stream to the \CLI.
At the \CLI, results are stored as a single file.

\subsubsection{Features}
Below, we list the features of the new system and link them to the requirements from \S\ref{requirement}.

\vspace{0.25em}\noindent
\textbf{Synchronized probing} -- A key limitation of the approach in~\cite{manycast} is that it performed measurements sequentially from individual VPs, 
leading to a significant time interval between outgoing probes to the same target.
To address this limitation, we use synchronized probing, where the \Server instructs \Clients to probe in parallel. 
\begin{itemize}[itemsep=1pt,topsep=1pt,labelindent=0.25em,leftmargin=*]
	\item Route changes in-between probes are less likely to occur when probing in parallel, since each target receives probes in a short timeframe, reducing the number of FPs (\ref{r:accuracy}).
	\item Reducing FPs from route changes (variable between subsequent censuses) improves longitudinal result stability (\ref{r:precision}).
	\item Fewer FPs translate to fewer targets to validate with a follow-up GCD-based measurement (\ref{r:responsibility}). 
	\item Measurements are performed by all \Clients in parallel, rather than sequentially, thus improving scalability (\ref{r:efficiency}).
	\item Probe offsets prevent bursts of probes to targets (\ref{r:responsibility}).
	\item Rate-limiting is less likely when spacing out probes (\ref{r:accuracy}).
\end{itemize}


\vspace{0.25em}\noindent
\textbf{IPv6 support} - The new system supports IPv6 probing.
\begin{itemize}[itemsep=1pt,topsep=1pt,labelindent=0.25em,leftmargin=*]
	\item Coverage is expanded with IPv6 censuses (\ref{r:coverage}).
\end{itemize}

\vspace{0.25em}\noindent
\textbf{Additional protocols} -- Our system supports ICMP, TCP, and DNS over UDP probing.
TCP probing uses \texttt{SYN/ACK} packets to high port numbers, for which we receive \texttt{RST} packets.
Since anycast is widely used for DNS infrastructures, we implemented
DNS probing with \texttt{A} and \texttt{CHAOS} \texttt{TXT} record queries.
\begin{itemize}[itemsep=1pt,topsep=1pt,labelindent=0.25em,leftmargin=*]
	\item Coverage is increased by supporting DNS and TCP probing (\ref{r:coverage}).
	\item TCP probing is done responsibly using \texttt{SYN/ACK} packets, which does not create state at the target host (\ref{r:responsibility}).
\end{itemize}

\vspace{0.25em}\noindent
\textbf{Failure awareness} -- When a \Client disconnects during a measurement (\eg, due to outside factors like link failure) 
the measurement is completed by the \Server using the remaining \Clients.
Disconnected \Clients automatically reconnect. 
\begin{itemize}[itemsep=1pt,topsep=1pt,labelindent=0.25em,leftmargin=*]
	 \item Outage at an anycast site will be handled by continuing measurements without this \Client (\ref{r:efficiency}).
\end{itemize}

\vspace{0.25em}\noindent
\textbf{Buffering and aggregation} -- At the start of a measurement, the \CLI sends the hitlist to the \Server, which streams it to the \Clients at the \CLI-defined probing rate.
Results captured at \Clients are forwarded to the \Server for aggregation.
\begin{itemize}[itemsep=1pt,topsep=1pt,labelindent=0.25em,leftmargin=*]
	\item Results are immediately submitted by \Clients, reducing the impact of \Client outages on results (\ref{r:efficiency}).
	\item \Clients do not store the hitlist nor results (\ref{r:efficiency}).
\end{itemize}

\vspace{0.25em}\noindent
\textbf{Deployment} -- Our anycast-based probing system can be deployed using a statically-linked binary (8\,MB) on both Linux and FreeBSD operating systems or Docker (image size of $\sim$8\,MB).
\begin{itemize}[itemsep=1pt,topsep=1pt,labelindent=0.25em,leftmargin=*]
	\item Docker allows for easy, OS-independent deployment (\ref{r:reproducibility}).
	\item A small image and binary size result in easy distribution and low storage requirements (\ref{r:reproducibility}).
\end{itemize}

\vspace{0.25em}\noindent

The source code is written in Rust and is publicly available with a
permissive license~\cite{mpl20} with extensive
documentation~\cite{gittooling}.

\subsection{Enumeration and Geolocation Pipeline}
\label{pipeline}
Figure~\ref{fig:pipeline} shows the measurement pipeline for our daily
anycast census.
Given the low probing cost and feasibility, 
we perform the
anycast-based measurements using TANGLED
(\S\ref{candidates}) toward full hitlists (\S\ref{hitlists}).
This measurement leads to a set of candidate anycast prefixes, which
we call \emph{anycast candidates} (\AC).
We extend these \ACs to include anycast prefixes found with GCD in
prior measurements (purple arrow)
allowing our census to also cover cases
where the anycast-based methodology fails to detect anycast
(\ie, FNs of \manycasttwo).
The initial set of prefixes, fed back into the measurement, is found
using large-scale GCD measurements and operator ground truth, which we
discuss later (\S\ref{subsec:gcdark}).

We perform GCD measurements towards
the \ACs.
For latency-based GCD measurements we need a large number of
geographically distributed VPs. 
The first two months of the census, GCD measurements were conducted
using TANGLED (32 VPs).
Since Jun.\ '24, we perform these measurements using CAIDA's Ark~\cite{ark},
which initially increased the number of VPs to $\sim$160 for ICMPv4 and $\sim$90 for
ICMPv6.
During the census, Ark has grown substantially as we 
use $\sim$250 ICMPv4 and $\sim$150 ICMPv6 VPs in Sep.\ '25.

We chose Ark due to its reliability, well-documented VP locations, and
geographical coverage.
Furthermore, Ark provides a Python API allowing for efficient collection of
results~\cite{scamper_api}.
GCD measurements allow us to enumerate and geolocate anycast sites using
iGreedy's analysis approach,
for which we make an improved version publicly available that severely reduces processing time~\cite{migreedy}. 
By performing the GCD measurement only on \ACs we
significantly reduce the probing cost by two orders of magnitude, from
millions to tens-of-thousands.
This reduction is necessary as measuring the entire hitlist with GCD
is impossible for a daily census at responsible probing rates that do
not overburden the Ark platform. 



The component that performs
anycast-based measurements used ICMP, TCP and DNS
for both IPv4 and IPv6.
We performed GCD measurements using ICMP and TCP.
We did not use DNS
due to the possible jitter introduced by DNS request processing by the
target that may inflate captured latency and affect the detection algorithm.
We have a monitoring system in place that warns when an upload fails,
when few VPs participate, or results deviate from the baseline.

\subsection{Census Output}

After this step we have two groups of results that are published in the census.
First, we have $\mathscr{G}$ which are \ACs confirmed with the follow-up GCD measurement.
Secondly, we use $\mathscr{M}$ to describe \ACs that are not confirmed with GCD.
Our daily census is finally constructed as follows.
For each prefix where either the anycast-based ($\mathscr{M}$) or GCD methodology ($\mathscr{G}$) detects anycast,
we publish the results to our public Git repository~\cite{gitcensus}.
The census includes information found using both methodologies for all protocols.
This data can be used by the community to evaluate anycast deployments with different levels of confidence.
Furthermore, we add the estimated number of anycast sites found by each measurement,
and for the latency-based measurements we include expected geolocations
generated using iGreedy's city population-based geo-detection algorithm.
We also include the number of VPs that participated,
which is variable since \textit{Ark} grows over time, as it impacts the accuracy of results
(see \S\ref{sec:longitudinal} for more details).


\section{Performance Assessment} \label{performance}
\begin{figure}[tb]
  \centering
  \includegraphics[width=\columnwidth]{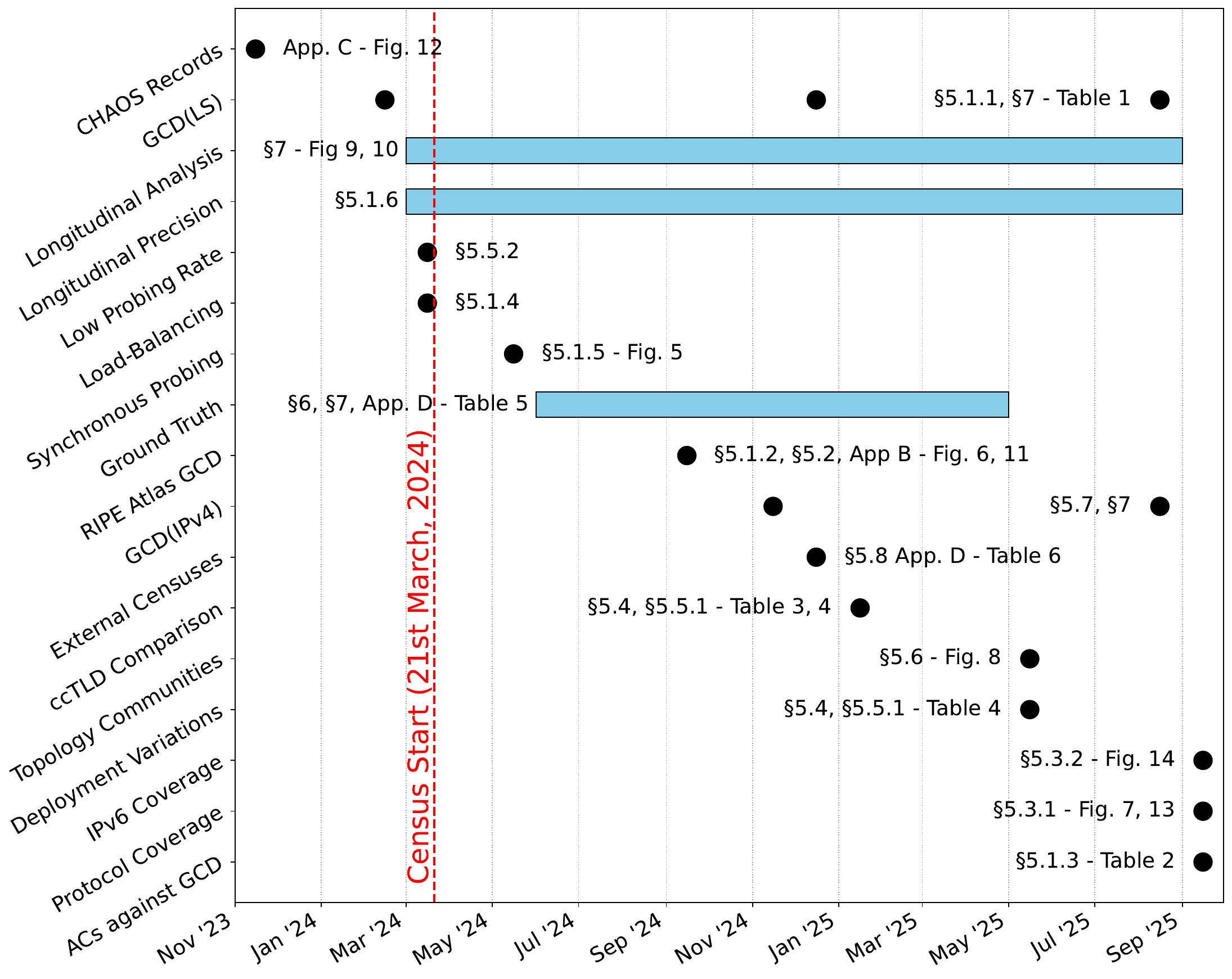}
  \caption{Roadmap of analysis conducted, detailing the date at which it was performed, the section in which we discuss the results, and the associated figures \& tables.}
  \label{fig:roadmap}
\end{figure}

In this section we assess the performance of \manycast.
We focus on: (\S\ref{subsec:accuracy}) accuracy and precision,
(\S\ref{sec:enumeration}) site enumeration,
(\S\ref{subsec:coverage}) benefits of broader protocol coverage,
(\S\ref{subsec:replicability}) replicability,
(\S\ref{subsec:responsible}) accuracy at low probing rates,
(\S\ref{subsec:topology}) impact of routing topology,
(\S\ref{sec:pfxsize}) prefix partitioning
and (\S\ref{subsec:external}) comparison against other censuses.
As we assess a longitudinal dataset,
we conduct analyses using data collected att different points of time,
summarized in Figure~\ref{fig:roadmap}.

\subsection{Accuracy and precision} \label{subsec:accuracy}

\begin{table}
\resizebox{\columnwidth}{!}{
\begin{tabular}{|c|r|r|r|r|r|}
    \hline
    \textbf{Protocol} & \multicolumn{1}{c|}{\textbf{\AC}}& \multicolumn{1}{c|}{\textbf{\gcdark}} & \( \textbf{\AC} \cap \textbf{\gcdark} \) & \textbf{FNs (FNR\%)} & \multicolumn{1}{c|}{\( \neg \textbf{\gcdark} \)} \\
    \hline
    \hline
    \textbf{ICMPv4} & 30,609 & 13,514 & 12,698 (94.0\%) & 816 (6.0\%) & 17,911 \\
    \hline
    \textbf{ICMPv6} & 11,166 & 11,286 & 10,617 (94.1\%) & 669 (5.9\%) & 549 \\
    \hline
\end{tabular}
}
\vspace{0.1em}
\caption{
	Comparing anycast candidates (\AC), excluding candidates from the feedback loop, to \gcdark.
}
\vspace{-2em}
\label{table:full-scan}
\end{table}

\subsubsection{Large-scale GCD measurement} \label{subsec:gcdark} 
As discussed in \S\ref{background}, GCD is highly accurate in detecting anycast.
However, the probing cost for GCD measurements on entire hitlists is too high for frequent measurements.
Because of the accuracy benefits of such a measurement, 
we perform periodic GCD measurements 
on the entirety of our IPv4 and IPv6 hitlists.
To limit impact, we performed the run at a low probing rate of 100 packets per second over a period of several days.
For clarity, we refer to this measurement as \gcdark
\footnote{Note: the \manycasttwo paper also discusses a GCD measurement, limited to a 2\% sample of prefixes to save on probing cost (\S4.4 of~\cite{manycast}).}.
This measurement provides a lower bound of the anycast deployments on the Internet,
since using the GCD approach can underestimate the number of anycast deployments (\S\ref{igreedy}).
We conducted \gcdark measurements in Feb.\ '24, Dec.\ '24, and Aug.\ '25 finding 13,684, 13,692, 13,514 /24s respectively.
Table~\ref{table:full-scan} shows the results for \gcdark from Aug.\ '25, compared to the ACs found using the anycast-based approach of \manycast (``\AC'' in Figure~\ref{fig:pipeline}).

The \gcdark ICMPv4 measurement inferred 13,514 anycast prefixes, of which 12,698 (94.0\%) are also inferred
by the anycast-based measurement in the \manycast pipeline.
Ergo, the anycast-based approach missed 816 (6.0\%) prefixes.
These prefixes are included using the feedback loop, thus providing coverage of these anycast-based FNs.
In \S\ref{sec:longitudinal} we discuss the longitudinal validity of the feedback loop.

As for accuracy, the anycast-based approach finds 30,609 ACs, of which 17,911 (58.5\%) were classified as unicast by \gcdark.
While these are mostly FPs of the anycast-based approach, ground truth shows these also contain TPs that we discuss in \S\ref{groundtruth}.
For IPv6 we find 11,286 anycasted /48s using \gcdark and 11,166 using \manycast and an intersection of 10,617 /48s.
This analysis shows there is value in doing large-scale latency measurements,
as this covers prefixes missed by the anycast-based approach.
However, there are drawbacks in terms of probing cost, timing, and infrastructure burden.
To balance the benefits, we repeat large-scale GCD measurements periodically
and feed the prefixes found into our \AC list so they are covered in subsequent daily census results.

\subsubsection{RIPE Atlas GCD measurement campaign} \label{subsec:ripe} 
To assess whether the daily pipeline provides sufficient geographical
coverage for GCD measurements, we ran an extensive measurement
campaign using 481 RIPE Atlas nodes in Sep.\ '24.
We selected nodes with stable uptime, 
user-reported location consistent with both MaxMind~\cite{maxmind} and IP2location~\cite{ip2location}
and ensured that no two nodes are within 100\,km of each other.
We conducted the measurement towards 23,821 ICMPv4 \ACs found on the 16th of September,
2024. 

Comparing the results,
we detect 12,186 (51\%) prefixes as anycast in the \manycast GCD measurement ($\mathscr{G}$), using 164 Ark VPs at the time,
and 12,202 (51\%) prefixes using RIPE Atlas, with an intersection of 11,953 prefixes.
RIPE Atlas missed 233 prefixes confirmed using GCD in the \manycast census, mostly due to probe measurement failures.
With RIPE Atlas we find 248 prefixes not found using GCD in the census.
Of these, 189 belong to Imperva, a CDN that offers DDoS protection.
We suspect these to be cases of temporary anycast which we discuss in \S\ref{sec:longitudinal}.
For the remaining 59 prefixes we see cases of regional anycast in areas where \manycast lacks VPs.

Overall these results show that RIPE Atlas provides relatively small gains in identifying anycast prefixes, and
in enumeration of sites as we show in \S\ref{sec:enumeration}.
Yet, even with increased spending quotas the measurement lasted three days
which is why we find it challenging to use RIPE Atlas in a daily \manycast pipeline.
We provide more detail in Appendix~\ref{app:gcd-atlas}.

\subsubsection{Investigating Disagreement}\label{subsec:disagreement} 

\begin{table}
\footnotesize
\begin{tabular}{|c|r|r|r|r|}
    \hline
     \textbf{\# of sites} & \textbf{Candidate} & \multicolumn{1}{c|}{ $\mathscr{G}$ } & \multicolumn{1}{c|}{ $\mathscr{M}$ } & \multicolumn{1}{c|}{\textbf{Overlap}}\\
     \multicolumn{1}{|c|}{\textbf{receiving}} & \multicolumn{1}{c|}{\textbf{anycast}} & \textbf{(\textbf{GCD})} & \textbf{(\( \neg \textbf{GCD} \))} &  \multicolumn{1}{c|}{\textbf{(in \%)}}\\
    \hline
\textbf{2} & 17,996 & 726 & 17,270 & 4.03\% \\
    \hline
\textbf{3} & 597 & 366 & 231 & 61.31\% \\
    \hline
\textbf{4} & 363 & 309 & 54 & 85.12\% \\
    \hline
\textbf{5} & 253 & 215 & 38 & 84.98\% \\
    \hline
\textbf{6-10} & 1,103 & 954 & 149 & 86.49\% \\
    \hline
\textbf{11-15} & 854 & 713 & 141 & 83.49\% \\
    \hline
\textbf{16-20} & 4,608 & 4,608 & 0 & 100.0\% \\
    \hline
\textbf{21-25} & 1,415 & 1,404 & 11 & 99.22\% \\
    \hline
\textbf{26-32} & 3,420 & 3,420 & 0 & 100.0\% \\
    \hline
        \hline
\textbf{Total} & 30,609 & 12,715 & 17,894 & 41.54\% \\
    \hline
\end{tabular}
\vspace{0.1em}
\caption{
	Comparing anycast-based ICMPv4 results per number of sites receiving responses with GCD.
}
\vspace{-2em}
\label{table:fp-breakdown}
\end{table}

The number of VPs receiving responses for a given target determines whether the anycast-based approach classifies a target as anycast.
For a VP count of one, the target is marked as unicast.
For a count above one, the target is marked as an anycast candidate.
The \manycasttwo~ paper showed the FP case (unicast responding to multiple sites) most commonly occurs when two sites receive responses.
Table~\ref{table:fp-breakdown} compares `GCD-confirmed' and `anycast-based only' (\ie, $\mathscr{G}$ and $\mathscr{M}$ as shown in Figure~\ref{fig:pipeline}),
based on the number of sites receiving responses.
%
We confirm that the vast majority of disagreement occurs when 2 VPs receive responses,
followed by fewer cases for 3, 4, and 5 VPs receiving responses.

When 4 or more VPs receive responses they are predominantly confirmed using GCD ($\mathscr{G}$), 
with only 393 of 12,613 prefixes not confirmed of which 87\% originate from Imperva.
We suspect this is related to their on-demand DDoS mitigation service, where prefixes are anycast for short periods of time.

Investigating the total prefixes not confirmed by GCD ($\mathscr{M}$), mostly received at 2 and 3 VPs, 
we observe 12,843 (> 70\% on any given day) originate from Microsoft's AS\,8075.
While we lack information to confirm this, we speculate, based on a 2019 study~\cite{todd-bgp-hard-2019},
that this is due to Microsoft's internal routing policies, where prefixes are announced globally to ingress traffic into their network as soon as possible,
with internal routing to a unicast destination.
We further suspect that replies egress at multiple edges, based on observed ingress routes, that reach different VPs.
Knowing of globally announced prefixes (\ie, global BGP) that route to a single location (\ie, unicast)
is valuable as it can help with understanding routing properties for such networks.
Using traceroute we confirm probes ingressing at distinct PoPs.
Future work is to include global BGP in our census and expand discovery of this practice.

\subsubsection{Influence of load balancers}\label{subsec:routing}
The \manycasttwo~ paper (\cite{manycast}, \S5.1) indicated that load balancers might be a cause of FPs when splitting traffic over links.
Load balancers typically calculate a hash using packet headers,
to determine the outgoing link.
Literature suggests load balancer hashes are mostly calculated using flow headers~\cite{mca, anycast_lb, LBpp}, which we keep static.
However, per-packet load balancers may still cause unwanted FPs.
To test whether our approach -- that varies the ICMP payload and checksum -- triggers 
load balancer decisions,
we performed the anycast-based measurement using static probes (\ie, all \Clients send exactly the same ICMP Echo Request, without payload and checksum variation).
The results match our regular measurement indicating most load balancers do not affect our results as we keep flow headers static.

\subsubsection{Benefits of synchronous probing}\label{subsec:sync}
The prototype implementation of \manycasttwo probed an entire hitlist from each VP in sequence, leading to 13-minute gaps between probes to the same target.
In \manycast we implemented synchronous probing, where \Clients send probes to a shared target with one second intervals from each other. 

\begin{figure}[t]
  \centering
  \includegraphics[width=\columnwidth]{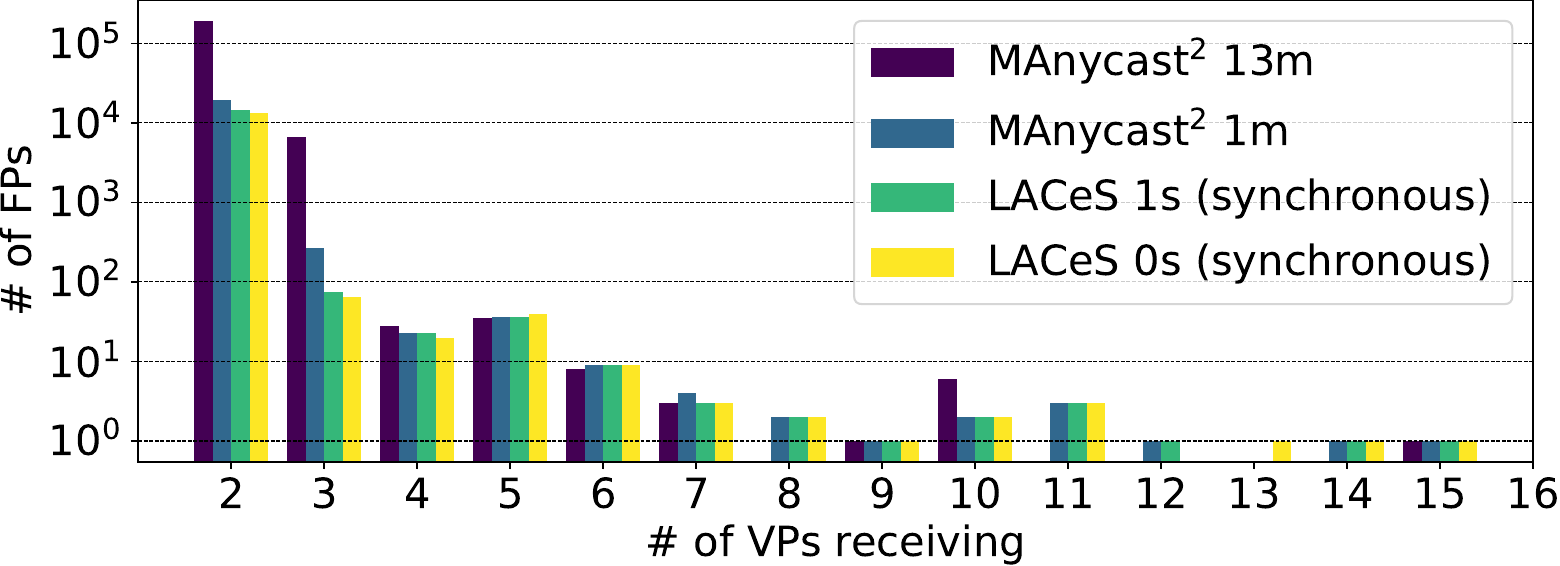}
  \caption{
  FPs found by number of VPs receiving using \manycasttwo~ with a 13- and 1-minute interval between probes, and \manycast with a 1- and 0-second interval between probes.
}
  \label{fig:fp-comparison}
\end{figure}

To compare our approach to the one used in the \manycasttwo paper, we performed a measurement using a 13-minute interval between probes.
This immediately demonstrates a shortcoming of the \manycasttwo approach: as our deployment has 32 sites, a single run takes around 7~hours.
We therefore also ran a second measurement using a 1-minute interval between probes, to obtain a shorter runtime.
We compare this to \manycast measurements with inter-probe offsets of 1~second and 0~seconds.
Figure~\ref{fig:fp-comparison} shows the the number of prefixes not confirmed using GCD ($\mathscr{M}$), which are mostly FPs, on a logarithmic y-axis against the number of VPs receiving on the x-axis.
We show FPs up to 16 VPs receiving as FPs are negligible when more VPs receive replies (see Table~\ref{table:fp-breakdown}).
Overall, the number of FPs increases as the inter-probe interval increases.
Probing with a 0-second interval yields 13,312 FPs, followed by 14,506 FPs for a 1-second interval, 19,830 FPs for 1~minute, and 198,079 FPs for 13~minutes.

An increased probing interval increases the time between first and
last probes to a target, leaving more time for route changes to occur.
We can therefore likely attribute the rise in FPs for larger intervals
to routing instability, causing a misclassification as anycast as stated in
the \manycasttwo paper~\cite{manycast}.
While reducing the inter-probe interval from 1~to 0~seconds leads to a
reduction of 1,194~FPs, we argue the benefits, in terms of responsible
probing, when spacing probes (like a regular ping) outweigh the slight
decrease in FPs.

\subsubsection{Longitudinal precision}\label{long-precision}
We assess the precision of our census by evaluating daily results from census start (March~21, 2024) to date (September~6, 2025) covering 17+ months,
focusing on prefixes detected using ICMPv4 for the anycast-based and GCD approaches.

On average, we observed 25.2\,k prefixes with the anycast-based approach and 12.9\,k prefixes with GCD daily.
Taking the union of all prefixes observed with the anycast-based approach during the 534-day timeframe, we find 203\,k /24-prefixes.
Of these prefixes, 9.9\,k (5\%) /24s are consistently observed on all 534~days
and 193\,k (95\%) prefixes only showed up on some days.
For the latter, we observe they are mostly not detected as anycast by GCD, and suspect many are FPs.
For the GCD-confirmed ($\mathscr{G}$) prefixes we observe 16.6\,k prefixes detected as anycast for at least one day, 9.7\,k (58\%) are observed every day.
We show in \S\ref{sec:longitudinal} -- where we perform a longitudinal analysis -- that the combined results of \manycast not only yield a stable census, but also captures both long-term ground-truth confirmed anycast deployments and captures anycast dynamics.

Overall, we observe that our anycast-based set has a high variability,
whereas the GCD set is much more stable,
confirming that our combined approach -- using anycast and GCD in \manycast -- is important to achieve high precision.

\subsection{Anycast site enumeration} \label{sec:enumeration}

\begin{figure}[t]
  \centering
  \includegraphics[width=\columnwidth]{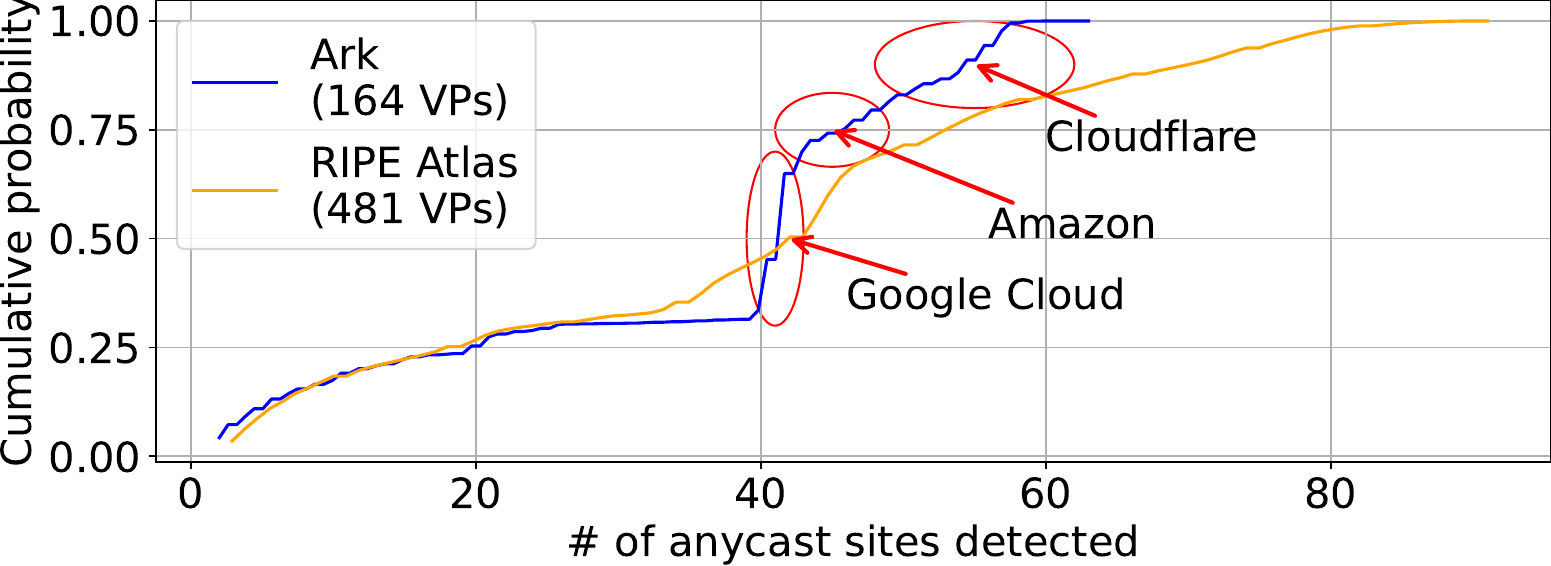}
  \caption{Number of sites detected per /24-prefix.}

  \label{fig:ark-counts}
\end{figure}

One feature from iGreedy that we include in \manycast is estimating the number of sites for an anycasted prefix.
Figure~\ref{fig:ark-counts} shows a cumulative distribution plot of the number of sites detected using latency data obtained from Ark (blue) and RIPE Atlas (orange) in Sep.\ '24.
At the time, Ark measurements were performed using 164 VPs.
The circles show the enumeration counts attributed almost entirely to prefixes from Google, Amazon and Cloudflare.
%
These counts are a lower bound of the actual number of anycast sites.
For example, Google has presence in 103 cities~\cite{google_locations} where we find $\sim$41 sites
and Cloudflare has PoPs in 300+ cities~\cite{cloudflare_locations} where we detect $\sim$54 sites.
For small deployments (< 30 PoPs) enumeration is near the actual number of sites which we validate using ground truth in \S\ref{groundtruth}.

Comparing RIPE Atlas with Ark,
we find both have similar results for small anycast deployments (bottom left).
For large deployments (top right),
RIPE Atlas is able to achieve a higher enumeration
of 80 compared to 60 with Ark.
However, it has large variability in the number of sites detected due to inconsistency in the number of RIPE Atlas nodes participating in measurements.
We provide a more elaborate comparison between these platforms in Appendix~\ref{app:gcd-atlas}.

Future work is to improve enumeration and geolocation data in our daily census, using, \eg, traceroute~\cite{chaos_traceroute}.
However, a complete enumeration for large anycast infrastructures is likely infeasible,
as also shown by Fan et al.\ that found 10\,k vantage points achieve an 80\% enumeration recall~\cite{chaos}.

\subsection{Increasing coverage}\label{subsec:coverage}

 \begin{figure}[t]
  \centering
  \includegraphics[width=0.7\columnwidth]{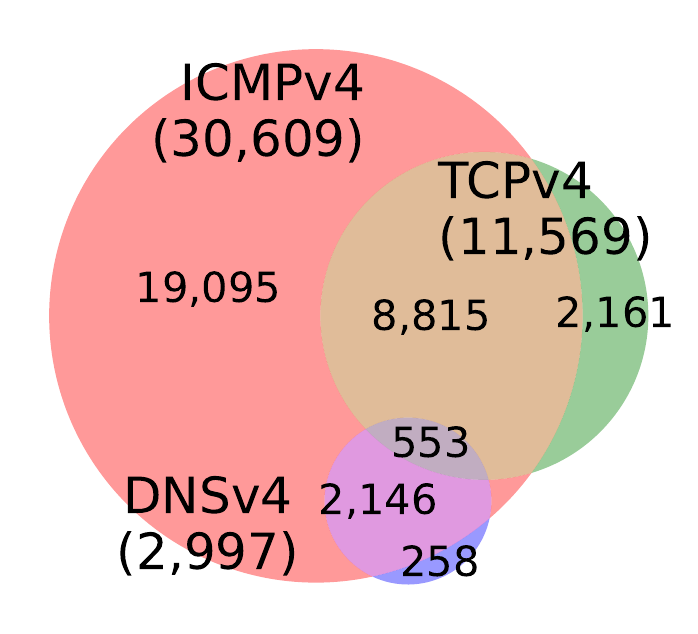}
  \caption{\manycast detection of anycast-based method for ICMPv4, TCPv4, and DNSv4.} 
  \label{fig:manycastv4}
  \vspace{-1.5em}
\end{figure}

\subsubsection{Protocol support}\label{subsec:protocols}
To extend coverage compared to iGreedy and \manycasttwo we added TCP and DNS over UDP probing.
Figure~\ref{fig:manycastv4} shows a breakdown of which prefixes can be detected using our anycast-based approach with different protocols.
We detect 30,609 anycast prefixes with ICMP,
11,569 with TCP, and 2,997 with DNS.

The largest group are ICMP-only prefixes, where we find 19,095 prefixes (57.7\%) (\ie, TCP and DNS did not detect these prefixes as anycast).
Next, we have the prefixes detected as anycast by both ICMP and TCP, totaling to 8,815, etc.

These results show that ICMP probing uncovers the most anycasted prefixes,
Furthermore, we see an increase in coverage using TCP and DNS, with 2,161 prefixes that are only detected as anycast using TCP,
and 258 with only DNS.
For the latter, we find 51 prefixes received at 4 or more VPs where the anycast-based approach has high accuracy as shown in Table~\ref{table:fp-breakdown}.
\textit{E.g.}, DNS G-root (operated by the US Department of Defense), LACNIC, Oracle, eBay, several registrars, and various TLDs deploy DNS anycast that is only detectable with anycast-based DNS probing.
For the 2,161 prefixes only visible with TCP, we confirm 167 with GCD. 

Prior work (see \S\ref{related_work}) also attempted to detect and enumerate anycast using \texttt{CHAOS} queries~\cite{chaos_traceroute}.
We therefore add support for \texttt{CHAOS} queries.
However, like prior work
we find \texttt{CHAOS} records are often used for co-located servers at a single location (suggested by values such as `auth1' and `auth2')
making it a weak indicator of anycast.
We provide a more detailed comparison to \texttt{CHAOS} records in Appendix~\ref{app:chaos}.

\subsubsection{IPv6 support}\label{subsec:ipv6}
Overall, we find fewer prefixes with the anycast-based approach for IPv6, 12,223 compared to 33,068 for IPv4.
This is unsurprising as our IPv6 hitlist is much smaller.
Like IPv4, most anycast candidates are detected using ICMP (11,840).
Furthermore, we find a large number of anycasted /48s responsive to TCP (8,208).
The higher TCP responsiveness, compared to IPv4, was expected due to the different hitlist origins.
Unlike the ISI hitlist, which is based on ping-scanning, TUM's and OpenINTEL's IPv6 hitlists more likely reflect targets running TCP services.
For the specific distribution for IPv6 we refer to Figure~\ref{fig:upsetv6} in the Appendix.

\subsection{Replicability} \label{subsec:replicability}

\begin{table}
\resizebox{\columnwidth}{!}{
\begin{tabular}{|c|r|r|r|}
    \hline
     \textbf{Protocol} & \textbf{Anycast candidates} & \multicolumn{1}{c|}{\textbf{Anycast candidates}} & \textbf{Intersection} \\
                               & \multicolumn{1}{c|}{\textit{our deployment}} & \textit{ccTLD deployment} & \\    
    \hline
    \textbf{ICMPv4} & 25,324 & 16,208 & 13,912 \\
    \hline
    \textbf{ICMPv6} & 6,996 & 6,501 & 6,255 \\
    \hline
\end{tabular}
}
\vspace{0.1em}
\caption{
	ICMP \ACs found using two distinct anycast deployments and their intersections.
}
\vspace{-2em}
\label{table:sidn}
\end{table}


To validate if our methodology is deployment-agnostic,
we deployed \manycast on an independent ccTLD registry-operated anycast production infrastructure providing VPs at 12 distinct locations.
Using this deployment we performed an anycast-based measurement in Jan.\ '25 for ICMPv4 and ICMPv6.
The results are shown in Table~\ref{table:sidn}.
We observe that the external anycast platform found fewer candidate anycast prefixes.
Especially for IPv4, they found 16,208 IPv4 anycasted /24-prefixes compared to 25,324 prefixes found with our own deployment.

In total, our census found 11,322 IPv4 anycast candidates not found by the ccTLD deployment.
For the non-intersecting \ACs, the vast majority (> 98\%) are detected by only 2~VPs.
Since the FP rate is high when 2 VPs receive responses,
this suggests the non-intersecting candidates are largely FPs.

Of 13,912 IPv4 \ACs seen on both platforms, 12,816 (92\%) are confirmed using GCD ($\mathscr{G}$).
Additionally, \ACs seen exclusively on our own or on the ccTLD platform include 481 and 112 GCD-confirmed prefixes, respectively.
These latter sets of prefixes are small and difficult to detect anycast prefixes only seen by either platform due to topological differences.
In \S~\ref{subsec:topology} we take a closer look at the impact of routing topology on detection.

For IPv6 we found 6,996 \ACs with our own and 6,501 with the ccTLD platform (intersection of 6,255).
Most non-inter\-secting \ACs are seen at 2 VPs and not GCD-confirmed.

We also performed anycast-based \manycast measurements using different deployments in May.\ '25
First, we deployed \manycast using an additional provider, \textit{Melbicom}~\cite{melbicom},
that provides 16 VPs
from which we announced our IPv4 prefix.
This deployment found 18,525 \ACs, covering 91.6\% of GCD-confirmed prefixes ($\mathscr{G}$) found that day.
Manual investigation of several missed GCD-confirmed prefixes shows they are mostly regional to Asia and/or Oceania
where \textit{Melbicom} provides only a single VP.
This indicates that \manycast requires multiple VPs in each continent to detect anycast confined within a continent.

To assess whether \manycast can be deployed using multiple hosting providers we announced our IPv4 prefix using both Vultr (as used for our daily census) and Melbicom
leading to 48 VPs combined.
In total, we find 31,340 \ACs covering 95.7\% of GCD-confirmed prefixes ($\mathscr{G}$) showing operators can deploy \manycast on diverse anycast deployments.
Surprisingly, we find more GCD-confirmed prefixes with the Vultr only deployment (96.2\%).

These results show that \manycast is deployment-agnostic and can also be deployed on anycast infrastructures that make use of multiple hosting providers.
We encourage operators to run their own \manycast census.
Additionally, we find detection of regional anycast is sensitive to topological connectivity of the deployment used.

A more detailed investigation of the effects of topology (\eg, upstream providers per PoP) may yield a better understanding as to when and why the FN and FP cases of the anycast-based approach occur.
In particular, we can use, \eg, reverse traceroute~\cite{reverse_traceroute} to find multi-path topologies that cause unicast targets responses to reach multiple sites.
\subsection{Reducing probing costs and impact}\label{subsec:responsible}

\subsubsection{Reducing deployment}\label{subsubsec:deployments}
Measuring the hitlist requires 188\,M probes for our daily ICMPv4 anycast-based measurement compared to the \gcdark from Dec.\ '24 requiring 1.3\,B probes.
For this reason, we only perform GCD towards \ACs found with the \manycasttwo approach.
The previous section showed
that a smaller anycast deployment finds considerably fewer \ACs, thereby reducing the number of targets for the follow-up GCD measurement.
However, this comes at a cost of FNs (GCD-confirmed anycast prefixes missed by the anycast-based approach).
This raises the question if we can reduce the anycast deployment used in the daily pipeline to reduce the number of \ACs with a minimal increase in FNs.

\begin{table}
\resizebox{\columnwidth}{!}{
\begin{tabular}{|lr|r|rr|r|}
    \hline
    \multicolumn{2}{|c|}{\textbf{Deployment}} & \multicolumn{1}{c|}{\textbf{\ACs}} & \( \neg \textbf{\gcdark} \) & (\( \neg \textbf{\gcdark} \) \%) & \textbf{Cost} \\
    \hline
    \textbf{\textit{EU-NA}} & 2 VPs & 12,492 & 2,164 & (15.8\%) & 12\,M \\
    \hline
    \textbf{\textit{1-per-continent}} & 6 VPs & 14,221 & 1,311 & (9.6\%) & 35\,M \\
    \hline
    \textbf{\textit{2-per-continent}} & 11 VPs & 27,379 & 633 & (4.6\%) & 65\,M \\
    \hline
    \textbf{ccTLD} & 12 VPs & 16,208 & 632 & (4.6\%) & 71\,M \\
    \hline
    \textbf{Melbicom} & 16 VPs & 18,525 & 1,155 & (8.4\%) & 94\,M \\
    \hline
    \textbf{TANGLED~\cite{tangled} (Vultr)} & 32 VPs & 25,396 & 524 & (3.8\%) & 189\,M \\ 
    \hline
    \textbf{Vultr+Melbicom} & 48 VPs & 31,340 & 593 & (4.3\%) & 283\,M \\
    \hline
    \hline
    \textbf{\gcdark Dec.\ '24 (full)} & 227 VPs & 13,692 & 0 & (0.0\%) & 1,335\,M \\
    \hline
\end{tabular}
}
\vspace{0.1em}
\caption{
	The number of anycast candidates (\ACs),
	missed \gcdark prefixes (\( \neg \textbf{\gcdark} \))
	and probing cost for various anycast deployments and \gcdark from Dec.\ '24 (full hitlist).
}
\vspace{-2em}
\label{table:reducing}
\end{table}

To answer this question we performed additional anycast-based measurements.
First, we pick two sites per continent, maximizing geographical distance (\eg, sites on the US East and West Coast), totaling 11 VPs (we have one site in Africa).
Then, we limit ourselves to the sites that receive the most responses on each continent (6~VPs).
Finally, we use only two VPs; one in North America and one in Europe.
Table~\ref{table:reducing} shows the result of these measurements, including the results of the ccTLD, Vultr (as used for TANGLED), Melbicom, and Vultr + Melbicom deployments.

Interestingly, we observe the second largest number of \ACs when probing with two sites in each continent.
This shows that reducing the number of VPs does not necessarily reduce the number of \ACs.
Despite the increase in \ACs, this deployment has a lower recall due to an increase of FNs compared to TANGLED. 
When probing with one site per continent, we find 14,221~\ACs.
However, this comes at a cost of 1,311~missed anycast prefixes.
Finally, when probing with only two VPs we find the fewest \ACs (12,492) at the cost of the most FNs (2,164).
This also shows that even with only two VPs, 84\% of \gcdark prefixes are detected,
which is unsurprising considering most anycast deployments have a global presence with at least one site in North America and Europe as we show in \S\ref{groundtruth}.
%
The FNs of each measurement largely consist of small anycast deployments that are confined to small geographical areas (\ie, regional anycast).

We argue it is impossible to determine an optimal number of sites, as the value of each VP highly depends on how it connects to the wider Internet.
We also argue that, despite a slightly higher probing cost, our current deployment has merit as it uncovers most difficult to detect regional anycast.

\subsubsection{Reducing probing rate}\label{subsec:low}
A key strength of \manycast is a low probing cost and hence a responsible measurement of anycast.
To illustrate this, we test another feature of \manycast.
We can configure \manycast to go over the hitlist at a specific rate, 
while maintaining one-second inter-worker spacings (\ie, time between \Clients sending their probe toward the same target).
To validate if we can maintain accuracy at a reduced probing rate, we perform a \manycast anycast-based measurement using a probing rate of 1/8th the `normal' rate of our daily census.
Even at this much-reduced rate, \manycast detects the same number of anycast candidates.
 
\subsection{Topology impact}\label{subsec:topology}
\begin{figure}[t]
  \centering
  \includegraphics[width=\columnwidth]{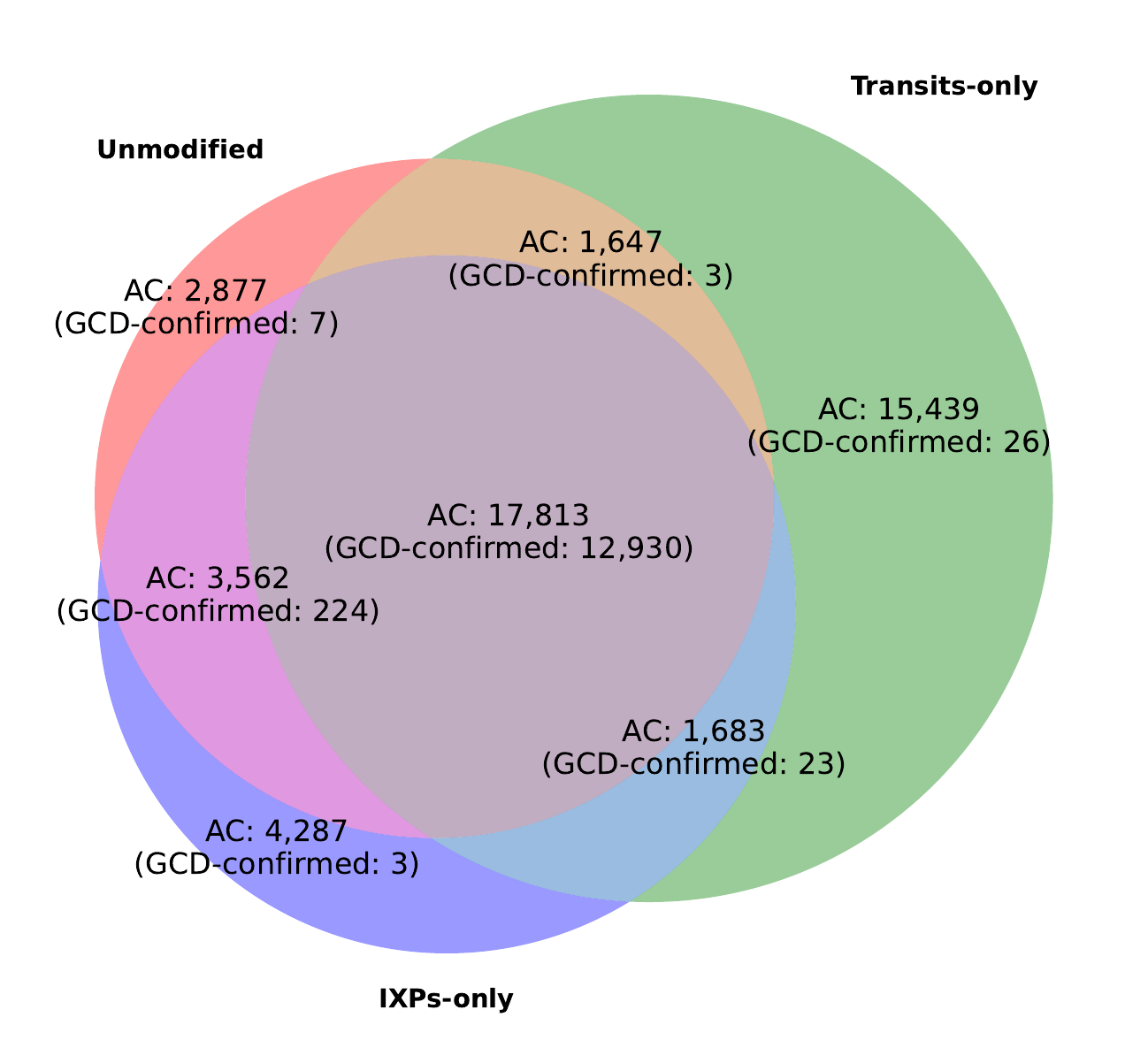}
  \caption{
  \ACs found using different routing policies.
}
  \label{fig:topology_venn}
\end{figure}

The previous section explored the effects of using different VPs and hosting providers,
showing the coverage of the anycast-based measurement depends on the number and locations of VPs used.
However, as noted in the \manycasttwo paper~\cite{manycast}, the number and diversity of upstreams at each location may have a large impact on coverage as well.
To further explore the effects of topological connectivity we change routing at each location using BGP communities supported by our upstream.
These measurements were performed in May.\ '25.
In particular, we use Vultr's \textit{Do not announce to IXP peers} and \textit{Announce to IXP route servers only}~\cite{vultr_communities}.
The former means our anycast prefix is only propagated to transit ASes (Transits-only) and the latter means our prefix is propagated only to IXP peers (IXPs-only),
thereby changing the path taken by reply traffic.
We provide a Venn diagram of the \ACs found using the various announcements in Figure~\ref{fig:topology_venn}.

Transits-only found the most \ACs (36,582) followed by IXPs-only (27,345) and the unmodified announcements found the fewest (25,899).
We suspect the Transit-only community found more \ACs as
transit ASes may have equal-cost paths to multiple PoPs,
where replies from unicast targets reach multiple VPs (\eg, due to ECMP).

Filtering on prefixes that are GCD-confirmed, we find that IXP-only found the most (13,033)
followed by the unmodified announcement (13,014)
and Transits-only found the fewest (12,713).
For these prefixes we find the anycast-based approach failed to detect them as candidate anycast using certain communities.
Surprisingly, the Transits-only community with the most \ACs had the fewest GCD-confirmed prefixes.
We speculate that some transit ASes may have a shortest or preferred path to a single VP causing anycast targets to reply to fewer VPs.
For instance, 1.1.1.0/24 has replies reach 31 VPs for both the unmodified and IXPs-only announcements whereas it reaches only 16 VPs for the Transits-only announcement.
 
These results give an indication as to how different routing topologies impact results.
Furthermore, it suggests that transit AS routing policies may have negative effects on the accuracy of the anycast-based approach.
 
\subsection{Anycast prefix size} \label{sec:pfxsize}
\vspace{0.1em}\noindent\textbf{BGP prefixes} --
As mentioned in \S\ref{methodology}, our methodology scans at a /24 (and /48 for IPv6) granularity as it is the
smallest prefix size propagated by BGP.
However, BGP announcements seen by route collectors are often less specific.
To assess whether such less specific prefixes are anycast in their entirety,
we look at all announced IPv4 prefixes in which we detect anycast.
This comparison is done using data from Apr.\ '24, where our census found 12,046 GCD-confirmed anycast /24-prefixes ($\mathscr{G}$).
Using CAIDA's prefix2as dataset~\cite{prefix2as}
we find those /24s to be part of 4,184 BGP announced prefixes.
Out of these BGP prefixes, 3,827 (91\%) are entirely anycast (\ie, each contained /24-prefix is anycast). 
For 70 prefixes (2\%) there is uncertainty as they contain at least one /24 that is unresponsive for the measurement.
The remaining 287 BGP prefixes (7\%)  contain unicast /24-prefixes according to our measurement.
This motivates our decision to scan at /24 granularity, rather than scanning at BGP announcement granularity which leads to overestimating anycast.

\vspace{0.1em}\noindent\textbf{Partial anycast} --
However, routing may yet be different within such prefixes~\cite{anycast_partitioning} .
%
This limitation was evident when examining inferences made with \textit{NTT DATA Global IP Network}, a large tier-1 transit provider,
who announce their address space at multiple PoPs (\ie, global BGP).
However, their addresses point to single servers (\ie, unicast).
The exception being 6 addresses (in a single /24) replicated at all PoPs for their public DNS resolver (\ie, anycast).
The census, which probed a unicast address for this prefix, classified the entire /24 as unicast.
We call such cases, where a /24 contains both unicast and anycast, partial anycast.

To assess the number of partial anycast prefixes, we perform a GCD measurement at /32 granularity targeting the entire allocated IPv4 address space, totaling four billion targets.
We performed this measurement in Nov.\ '24 and Aug.\ '25, which we refer to as \gcdfull.
For the Aug.\ '25 measurement we used 13 VPs spanning multiple continents over a period of ten days.
While GCD measurements typically require a large number of VPs for accurate detection,
we believe in this case a few VPs suffice for detecting partial anycast as partial anycast requires a private backbone with global reach
to allow for different routing topologies within an announced prefix.

This scan revealed anycast in 12.4k /24s, for which we inferred partial anycast in 1,025 (8\%) /24s (including the aforementioned tier-1 prefix).
We include a partial anycast flag in our census to avoid overestimating anycast.
%
We observe most partial anycast prefixes are announced by large CDNs that have global presence.


Surprisingly, we found more (1,483) partial anycast prefixes in Nov.\ '24 of which 793 were not found in Aug.\ '25.
Inversely, 335 partial prefixes in Aug.\ '25 were not found in Nov.\ '24.
Investigating these non-intersecting prefixes, we observe they are mostly seen as entirely unicast within a short period.
Additionally, they are largely originated by Imperva which we suspect to be temporary anycast as observed in our RIPE Atlas measurements. 

This analysis suggests partial anycast may be short-lived in many cases.
While interesting to study, the high measurement cost (a full GCD of the IPv4 address space) and impact of such measurements make a longitudinal study of this phenomenon undesirable.
Our current plan is to repeat \gcdfull on an annual basis to capture the fact that such dynamics exist while accepting the limitation that we cannot capture the dynamics of this phenomenon in more detail
 
\subsection{Comparing with external sources} \label{subsec:external}

There are also external datasets that contain information on anycast.
We compare our work against two, a commercial dataset from \textit{IPInfo} and a public dataset from \textit{BGPTools} using census data from Dec.\ '24.
As our census has high confidence in GCD-confirmed prefixes ($\mathscr{G}$) we focus the comparison on these results.
 
\textbf{IPInfo}, a commercial provider of IP address databases, shared their list of detected anycast prefixes with us~\cite{ipinfo}.
First, for IPv6 our census detects anycast in 6.3k /48s whereas IPInfo detects anycast in 2.0k /48s.
For IPv4, our census detects anycast in 13.4k /24s and \textit{IPInfo} in 14.0k /24s.
Overall, 12.6k /24s are found in both datasets showing high agreement.
Inspecting the 0.8k prefixes found only in our census, we find most are detected in few locations and are regional.
For the 1.4k prefixes found only by \textit{IPInfo}, we observe 0.6k originate from Imperva and several hundreds from other CDNs offering anti-DDoS services
where we suspect temporary anycast.
We believe \textit{IPInfo} to include larger numbers of temporary anycast as they accumulate anycast prefix using weekly snapshots.

\textbf{BGPTools} uses an anycast-based approach similar to the first stage of \manycast.
However, there are two key differences: 1) if \textit{BGPTools} detects a single address in an announced prefix as anycast, they classify the entire prefix is anycast~\cite{bgptools_assumption}, and 2) \textit{BGPTools} do not use GCD to filter out FPs.
\textit{BGPTools} marks 3,047 announced prefixes as anycast, of those 3,047 prefixes our anycast-based stage marks 2,954 (97\%) as anycast.
Of these, 228 (8\%) prefixes are detected at two of our VPs, and our GCD stage marks these as FPs, suggesting \textit{BGPTools} overestimates the number of prefixes that use anycast.
Our census finds 13,495 anycast /24-prefixes confirmed by GCD ($\mathscr{G}$) in total for that day.
Of these, 9,739 (72\%) are also covered by the \textit{BGPTools} prefixes, leaving 3,756 (28\%) /24-prefixes they miss.

Investigating the prefixes \textit{BGPTools} classifies as anycast, we find 467 (15\%) are less-specific than a /24.
Comparing this to our census, we find 60 of these prefixes contain a grand total of 8,038 unicast /24s, illustrating the risk of classifying entire announced prefixes as anycast.

Looking at IPv6, \textit{BGPTools} detects 1,148 IPv6 prefixes,
1,131 of which are covered by our census while 8 are missed as they are not in our hitlist.
That same day we find 6,358 anycast /48s, 1,479 (23\%) of which were not found by \textit{BGPTools}.

These results reaffirm that IP anycast cannot be generalized to BGP prefixes, that our census has better coverage, and that our pipeline obtains a higher TP rate.


\section{Ground truth validation}\label{groundtruth}
Finally, we validate our GCD-confirmed results ($\mathscr{G}$) against public datasets and reach out to operators to obtain ground truth. 

 \begin{table}[t]
 \footnotesize
\begin{tabular}{|c|c|rc|rc|}
    \hline
     \textbf{AS} & \textbf{Organization} & \multicolumn{2}{c|}{\textbf{IPv4 (/24)}} & \multicolumn{2}{c|}{\textbf{IPv6 (/48)}} \\
    \hline
    \rowcolor{orange!30}
    \textbf{396982} & Google Cloud & 3,627 & (1st) &  \cellcolor{white!100} 5 & \cellcolor{white!100} \\
    \hline
    \rowcolor{green!30}
    \textbf{13335} & Cloudflare &  \cellcolor{green!50!black} \cellcolor{green!50!black} {\color{white}3,133} &  \cellcolor{green!50!black} \cellcolor{green!50!black} {\color{white}(2nd)} & 284 & (3rd) \\
    \hline
    \rowcolor{orange!30}
    \textbf{16509} & Amazon & 1,286 & (3rd) & \cellcolor{white!100} 120 & \cellcolor{white!100} \\
    \hline
    \rowcolor{green!30}
    \textbf{54113} & Fastly &  \cellcolor{green!50!black} {\color{white}435} &  \cellcolor{green!50!black} {\color{white}(4th)} & 65 & \\
    \hline
    \rowcolor{green!30}
    \textbf{209242} & Cloudflare Spectrum & \cellcolor{green!50!black} {\color{white}289} & \cellcolor{green!50!black} {\color{white}(5th)} & 3,338 & (1st) \\
    \hline
    \textbf{19551} & Incapsula (Imperva) & 2 & & 352 & (2nd) \\
    \hline
    \textbf{12041} & Afilias & 221 & & 222 & (4th) \\
    \hline
    \textbf{44273} & GoDaddy & 32 & & 122 & (5th) \\
    \hline
\end{tabular}
\vspace{0.1em}
\caption{
	Largest ASes by number of anycast /24-prefixes.
}
\vspace{-2em}
\label{table:largest_ases}
\end{table} 
 
\vspace{0.1em}\noindent\textbf{DNS operators} -- 
We find all IPv4 and IPv6 prefixes for the Quad9 public resolver, RIPE authoritative nameservers,
and all DNS root servers correctly classified as anycast by our census.
Interestingly, we observe that G-root is unresponsive to ICMP and TCP for both IPv4 and IPv6.
However, \manycast is able to detect it using DNS.

Next, we reached out to multiple ccTLD operators as they often deploy regional anycast that is difficult to detect.
These include \textit{.be}, \textit{.cz}, \textit{.de}, \textit{.dk}, \textit{.nl}, \textit{.nz}, \textit{.ua}.
The census is able to detect the vast majority of anycast nameservers. 
The exceptions are two anycast deployments regional to Belgium and the Netherlands not found with GCD, of which one is detected with the anycast-based approach,
3~nameservers regional to New Zealand, one nameserver local to Germany, and 2~IPv6 prefixes not covered by our hitlist.
Furthermore, we observe our GCD reported locations closely match reality,
exceptions being multiple sites in a single city or nearby cities (\eg, Prague, Bratislava, Vienna) being detected as a single site.
Finally, a few anycast operators expanded their deployment during the census which is visible in our longitudinal data.
 
\vspace{0.1em}\noindent\textbf{Hypergiants} -- 
 Table~\ref{table:largest_ases} shows the largest ASes based on the number of IPv4 /24 and IPv6 /48 anycast prefixes they originate.
We find that Google Cloud is leading in IPv4 anycast
followed by Cloudflare, Amazon, and Fastly.
These ASes are large CDNs, often referred to as hypergiants.
We color the table based on ground truth validation;
green for operator-confirmed ground truth (no FPs),
dark-green when fully accurate (no FPs or FNs),
and orange when confirmed with public data.
Due to the dominance of hypergiants, these results comprise 59\% of our IPv4 and 63\% of our IPv6 census.

Google provides an \textit{ipranges} dataset~\cite{google}
that discloses where prefix ranges are announced for Google Cloud resources.
We compared our census to this dataset for January~6, 2025.
Google's dataset contains 33 globally announced IPv4 prefixes ranging from a /16 to a /24, totaling 3,581 /24-prefixes.
Out of these prefixes, 3,572 are detected as anycast in our census.
For the 9 missing prefixes, 8 are not in our hitlist
and 1 we detect as unicast (which we confirm to be unicast using traceroute).
The 55 remaining Google Cloud prefixes we detect as anycast are not listed in the \textit{ipranges} dataset.

Cloudflare, the CDN that ranks first for IPv6 and second for IPv4 anycast in the census, shared ground truth data with us.
Their data confirms our census is fully accurate for IPv4 anycast, where we have no FPs and no FNs.
For IPv6, we have no FPs,
however, our census misses IPv6 prefixes due to IPv6 hitlist limitations.

Next, we evaluate the prefixes we detect as anycast for Amazon
using an \textit{ip-ranges} dataset~\cite{amazon} similar to Google.
The list contains 37k~/24~prefixes listed as globally announced,
these are cases of global announcements but not necessarily anycast as discussed previously.
This list includes partitions more specific than /24, even /32 partitions.
For the 1,286~/24-prefixes we detect as anycast, 1,123 (87\%) are listed as being announced globally,
2 are listed as being announced in the east coast of the US,
 and 161 (13\%) are not listed.

We also reached out to Fastly whose prefixes feature prominently in our dataset and obtained ground truth data.
They disclosed a traffic engineering implementation that uses \textit{backing anycast},
where a large prefix is announced globally (anycast) and more specific prefixes within route to a single PoP (unicast),
which allows for withdrawing the unicast announcement to fall back on the less-specific anycast announcement.
Such traffic engineering allows for rapid redistribution of traffic,
\eg, if a unicast prefix announced at a single PoP is the victim of a DDoS attack.
For IPv4 Fastly confirmed all /24-prefixes we find are anycast.
However, for IPv6 we have a considerable number of FPs.
After troubleshooting, we discovered that two \textit{Ark} nodes are located in ASes
that miss /48 announcements for the CDN due to these ASes filtering these out or their upstream ASes not propagating the announcements.
In the case of these FPs the probed addresses are unicast using a /48 with a less specific \textit{backing anycast} prefix (the TE approach mentioned above).
Due to these nodes being unaware of the /48, they are routed to the nearest PoP, the address is then misclassified as anycast.
Filtering out these Ark nodes from our IPv6 results,
we find all detected IPv6 anycast to match ground truth.

These results indicate our methodology detects ASes that, \eg, do not propagate /48-prefixes,
by performing anycast-based measurements where we announce a specific prefix at a single site.
Finally, the TE practice used to switch prefixes from unicast to anycast when a PoP is overloaded motivates the need for a daily census.


    %

\begin{figure*}[t]
  \centering
  \includegraphics[width=\textwidth]{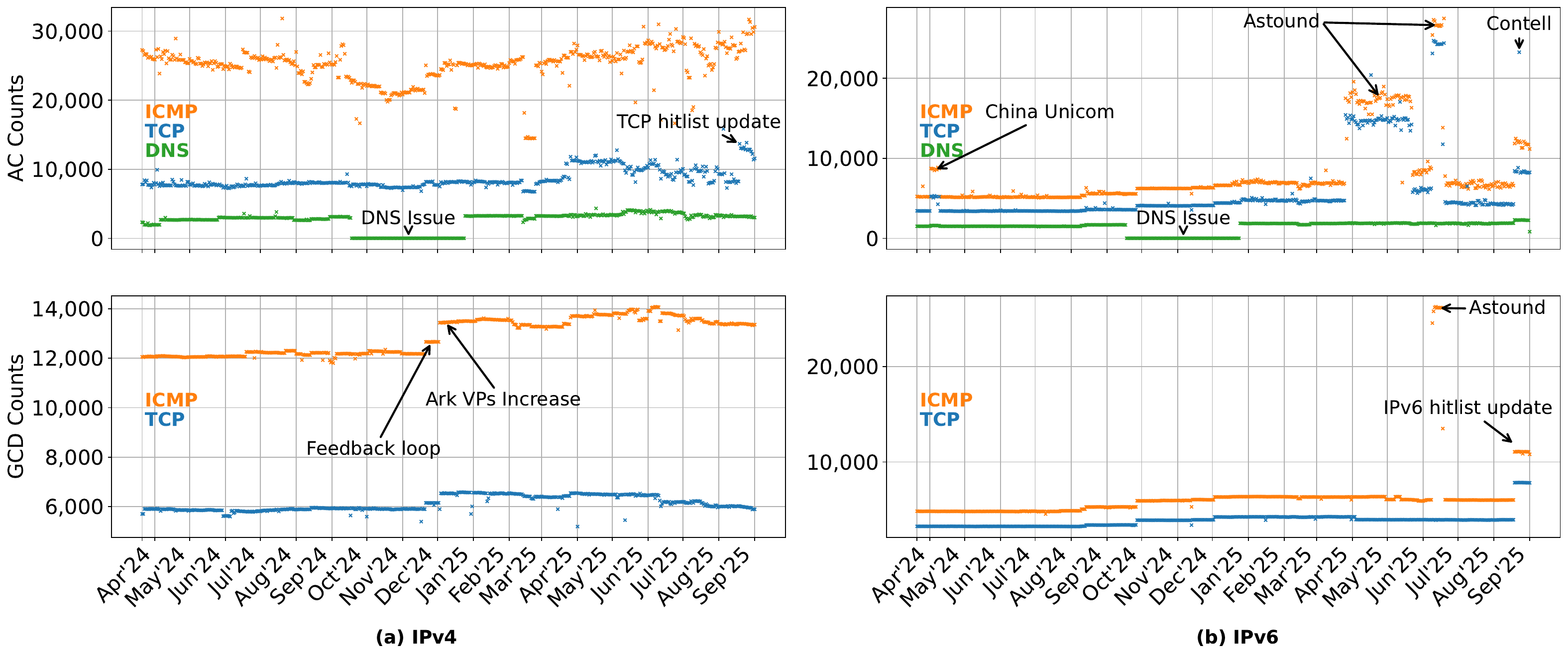}
    \caption{
    \manycast detection statistics by method and protocol over time for both IPv4 and IPv6.
}
  \label{fig:stats-lon}
\end{figure*}

\section{Longitudinal View}
\label{sec:longitudinal}

This section provides a longitudinal view of \manycast, covering daily census data between
Mar.\ '24 and Sep.\ '25 (17+ months).

\vspace{0.1em}\noindent\textbf{Hitlist and feedback loop} -- The coverage of our census is dependent on the hitlists used, which we update quarterly to maintain high coverage.
First, the USC/ISI hitlist~\cite{isi} updates quarterly due to churn in ping-responsive hosts.
However, we observe this churn mostly affects unicast hosts and has limited effects on \manycast.
The change in coverage for IPv6 can be explained by the growth of the hitlist~\cite{v6hitlist}.
%

Furthermore, prefixes found using \gcdark are fed into the feedback loop to cover rare cases of regional anycast undetectable by the anycast-based approach.
We conducted the \gcdark measurement in Feb.\ '24, Dec.\ '24, and Aug.\ '25 finding 13,684, 13,692, 13,514 /24s respectively.
Comparing those found in Feb.\ '24 and Aug.\ '25, we find an intersection of 11,719 /24s that remained anycast.
For 1,965 /24s found in Feb.\ '24, we observe they are no longer anycast and 1,796 new prefixes are detected as anycast in Aug.\ '25.
These results highlight the need for frequent \gcdark measurements.
%
For IPv6 the \gcdark sets had 4,891, 6,314, and 11,286 /48s respectively (on the same dates as IPv4).
Unlike IPv4, we see large variations as coverage is largely limited by the IPv6 hitlists used.

Similarly, we periodically repeat \gcdfull sweeps
to provide coverage for partial anycast and ping-responsive anycast prefixes missed by our hitlists.

\vspace{0.1em}\noindent\textbf{Daily ACs} -- 
To demonstrate the value of daily ACs from the anycast-based approach (instead of solely relying on prefixes found using \gcdark)
we compare Dec.\ '24 \gcdark prefixes with three snapshots in January, May, and August 2025.
We find that as time progresses our daily results deviate substantially from this one time snapshot.
We find 40, 255, 399 GCD-confirmed prefixes from daily \ACs on the respective snapshots that were not found with \gcdark in Dec.\ '24.
These results clearly show the value of daily \ACs to maintain coverage (in particular for short-lived anycast).

\vspace{0.1em}\noindent\textbf{Detection counts} --
Figure~\ref{fig:stats-lon} shows the number of prefixes found using all of our probing methods for both IPv4 and IPv6.
First, looking at the IPv4 statistics we observe variability for the number of ACs found using the anycast-based measurements.
This is due to routing dynamics causing false positives in the anycast-based approach as explained in \S\ref{subsec:disagreement}.
For GCD we observe an increase in Dec.\ '24 when we updated the feedback loop with \gcdark prefixes and Jan.\ '25 when the number of Ark VPs increased.


Looking at IPv6, we observe growth in \ACs and GCD-confirmed prefixes as our IPv6 hitlist grows.
Most notably in Aug.\ '25 where the number of GCD-confirmed prefixes doubled.
In Apr.\ '24, we see an uptick in IPv6 ACs belonging to China Unicom followed by large increases in April-June and July from Astound Broadband (a large US telco) and a peak in August from contell (a Russian retailer).
We suspect these ASes experienced routing instability affecting a large number of /48s due to large IPv6 prefix sizes.
Interestingly, we confirm the 20\,k /48s increase in \ACs from July using GCD.
Closer investigation shows these belong to two /29 IPv6 prefixes announced by Astound (visible in the public routing table),
which iGreedy locates to Baltimore and New York City.

 
\begin{figure}[t]
  \centering
  \includegraphics[width=\columnwidth]{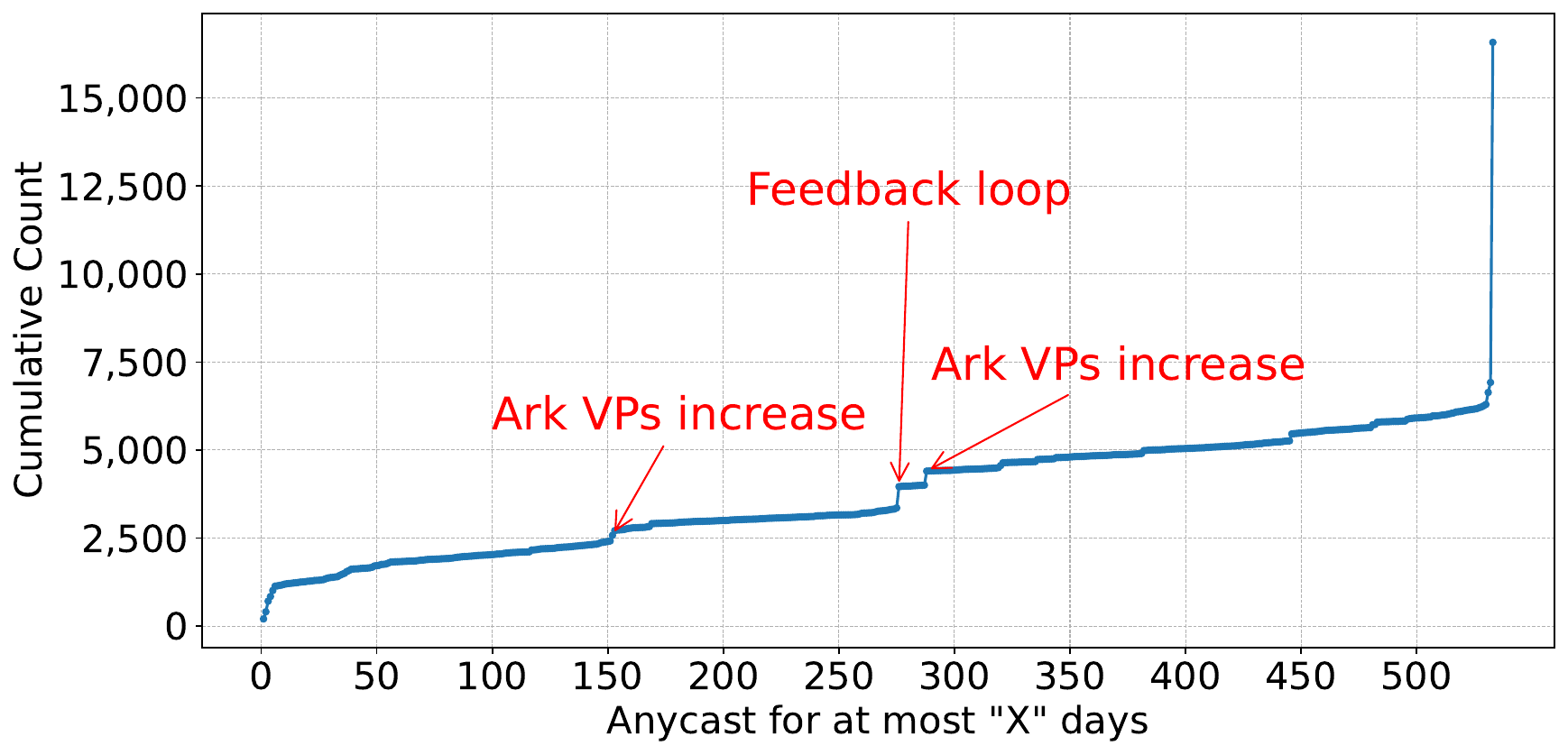}
  \caption{
  Cumulative count of anycast /24-prefixes detected to be anycast for at most ``X'' days.
}
  \label{fig:anycast-lon}
\end{figure}
 
\vspace{0.1em}\noindent\textbf{Persistent anycast} --
Figure~\ref{fig:anycast-lon} shows the cumulative counts of anycast prefixes found during the census period,
based on the number of days they were detected as anycast.
The first thing that stands out, on the right of the graph,
are 9.7\,k (60\%) /24s, belonging to 433 distinct ASes, that were anycast for the entire period.
Of these prefixes, 66\% were confirmed via ground truth in \S\ref{groundtruth};
3.2\,k (33\%) from Google, 
2.6\,k (27\%) from Cloudflare 
and 1.2\,k (12\%) from Amazon. 

We find 3\,k (19\%) prefixes observed to be anycast for only 20-80\% of the measured period,
603 of these belong to Google (AS\,396982) and 402 to Fastly (AS\,54113) where we suspect dynamic address utilization (\eg, \textit{backing anycast} as discussed in \S\ref{groundtruth}).
For the remainder there exist multiple reasons;
i) changes in \manycast coverage,
ii) prefixes that became anycast during the period,
iii) or prefixes that dynamically switch between unicast and anycast.
Cases ii) and iii) can only be observed through using the daily \ACs from the anycast-based measurement

%
The bottom left shows cases of temporary anycast. 
In total, we find 1,069 /24s that were anycast for a maximum of 10 days during our census period.
Of these, 572 belong to Imperva (AS\,19551) that offers content caching, and DDoS mitigation services for customers using their anycast network.
%
Finally, we find 191 prefixes that were anycast only for a single day.
Manual inspection shows these are mostly anycasted at two locations (\ie, the unicast location and an anomalous second location).
We suspect these are BGP misconfiguration or hijacking events, matching observations by Cicalese et al.\ that hypothesized iGreedy could detect BGP hijacking.
Future work is detection of such events using \manycast, including detection based on BGP data to cover short-lived events not visible with daily measurements.

\vspace{0.1em}\noindent\textbf{Enumeration} -- 
We observe enumeration for large anycast deployments increasing longitudinally as the Ark platform expands (\eg, for Cloudflare we find $\sim$30 sites in Mar.\ '24 compared to $\sim$70 sites in Sep.\ '25).
Furthermore, we observe deployment changes in our daily census, \eg, during our census we find additional locations for \textit{.cz} matching infrastructure expansions disclosed by the operator.

\vspace{0.1em}\noindent\textbf{Technical events} --
From 19th of September 2024 till 24th of December 2024, our census collected no DNS results due to issues with the tooling that was incorrectly flagging all replies as invalid.
Following this issue we added an alerting system that triggers when canary checks fail or results substantially deviate from the baseline. 

In Figure~\ref{fig:stats-lon} we observe large drops in ICMPv4 ACs, 
these are caused by \Clients losing connection with the \Server
resulting in anycast-based measurements with fewer VPs.
We fixed this issue in Jul.\ '25 by implementing automatic re-connects thus improving robustness of \manycast.
We do note this issue had minor effects in GCD-confirmed prefixes thanks to the feedback loop in \manycast.


\section{Discussion}\label{lessons}
\vspace{0.1em}\noindent\textbf{Lessons learned} --
We find the performance of both anycast-based and GCD measurements depend on the number, and geographical and topological diversity of sites.
Since the majority of anycast is from hypergiants with global deployments,
we observe that few nodes on different continents can detect the majority of anycast, as also exemplified by \textit{BGPTools}~\cite{anycatch}.
However, for detecting regional anycast a larger measurement platform is required.
Furthermore, enumeration and geolocation of anycast sites requires a well-distributed measurement platform.

We find our census provides good coverage of the anycast landscape,
as confirmed by extensive validation.
However, for IPv6 we are restricted by hitlist coverage.
Additionally, for enumeration we provide a lower bound
as latency-based measurements struggle to differentiate between anycast sites in near-proximity.

Whilst the \ACs found using the \gcdark scans (targeting the entire hitlist) can be used solely as input for the daily GCD scans (targeting only \ACs),
we find the daily anycast-based measurements (targeting the entire hitlist) allows us to cover prefixes that switch from unicast to anycast.
Additionally, we provide coverage of partial anycast prefixes, \ie, prefixes containing both anycast and unicast addresses, using \gcdfull scans.

When using our census, we advise to keep the FP limitation of the anycast-based approach in mind,
especially when two VPs receive responses.
As for GCD, we find it is highly accurate but has rare cases of FNs due to regional anycast.

We encourage operators to run their own census using the \manycast tool, which supports both anycast- and latency-based measurements.
Furthermore, our daily \AC lists are made available in our public census, making it possible to combine them with other \AC lists to improve coverage.

\vspace{0.1em}\noindent\textbf{Community value} --
Both operators, governments, and academics benefit from knowing which services are anycasted (and where they are located) to
perform risk resilience assessment, determine regulatory compliance, and much more.
Furthermore, by performing the census daily we track such issues longitudinally.
We foresee that the dataset may have values for studies that explore, \eg, Internet resilience or locations of Internet services.
We commit to continuously provide this service to the community in years to come and make the dataset publicly available.

\vspace{0.1em}\noindent\textbf{Limitations} --
\manycast has limitations that affect the coverage of the daily census we provide.
As we intend on providing this daily census for the foreseeable future, we address those limitations here and indicate future plans to resolve them.

\manycast only covers prefixes on hitlists containing hosts responsive to its probing methods.
We intend to add probing methods and contribute to improving hitlists.
%
For daily coverage provided by the anycast-based measurements,
we require an anycast deployment with topological diversity.
We are investigating the use of BGP communities and additional providers to
improve coverage and reduce false positives.
For GCD, accuracy is dependent on the number and distribution of VPs,
as Ark grows we expect improved coverage of regional anycast.
Finally, we intend on extending \manycast by performing GCD using DNS.
%


\section{Conclusions and Future Work} \label{conclusion}

We built LACeS, a new system that combines the \manycasttwo anycast measurement approach of Sommese et al.~\cite{manycast}
and the Great Circle Distance (GCD) approach pioneered by Cicalese et al.\ with iGreedy~\cite{igreedy} to build an accurate and efficient daily anycast census.
Through extensive validation we show that \manycast allows for an efficient, accurate and precise daily anycast census for both IPv4 and IPv6
and with support for transport layer probing using UDP/DNS and TCP.
We intend to provide daily \manycast censuses to the community for the foreseeable future, covering 17+months as of Sep.\ '25,
through a public Git repository~\cite{gitcensus}
and release the entire \manycast toolchain under a permissive open source license~\cite{gittooling}.

We intend to further extend \manycast by including a trigger-based detection of anycast not visible with daily census granularity, \eg, using BGP route collectors,
and provide an API to the community for live measurement of anycast.
Finally, we are planning to use \manycast to detect suspected BGP hijacking.

\begin{acks}
This work was supported by 
the RIPE NCC Community Fund,
INTERSCT (NWO grant NWA.1160.18.301),
the National Science Foundation (NSF grant OAC-2131987),
and the European Commission's G\'{E}ANT GN5-2 Project.
We thank the operators that shared ground truth data with us,
including
Wouter de Vries (Cloudflare),
Stephen Strowes (Fastly),
Colin Petrie (Global NTT),
Marco Davids (SIDN),
Josh Simpson (InternetNZ),
Carlos Collao Vilches (Punktum),
Thomas Dupas (DNS Belgium),
Jaromir Talir and Tom\'{a}\v{s} H\'{a}la (CZ.NIC).
We also thank Oliver Gasser (IPInfo) for sharing their detected anycast prefixes
and Thijs van den Hout (SIDN) for performing \manycast measurements using their ccTLD deployment.
Finally, we thank our shepherd for their guidance and all reviewers whose feedback helped improve this paper.
\end{acks}

\bibliographystyle{ACM-Reference-Format}
\balance
\bibliography{manycast}

\appendix
\newpage


\section{Ethics}
Our ethics review board does not require a specific ethics approval for measurements 
conducted according to community best practices. In this work, we follow these practices 
by ensuring that the address space of our measurement infrastructure has correct pointers 
to abuse contacts and references to a page explaining the goal of the measurements, with 
clear instruction for opting out.
The analysis conducted in this paper was also helpful
in establishing a significantly lower and manageable probing rate.
In the early phase of creating \manycast, we prioritized such analysis
 which allowed us to design \manycast to be responsible
 and minimize the performance impact on targeted infrastructure.
Figure~\ref{fig:roadmap} shows these early measurements at the top of the figure.

\section{GCD using RIPE Atlas} \label{app:gcd-atlas}
We performed GCD using RIPE Atlas for the 23,821 \ACs of 16~September, 2024.
This measurement lasted three days due to rate-limiting, even with increased spending quotas.
Furthermore, it incurred a cost of 37 million RIPE Atlas credits
which would take a volunteer hosting a RIPE Atlas probe approximately five years to accumulate.
For these reasons we find RIPE Atlas unsuitable for the daily census.

\begin{figure}[t]
  \centering
  \includegraphics[width=\columnwidth]{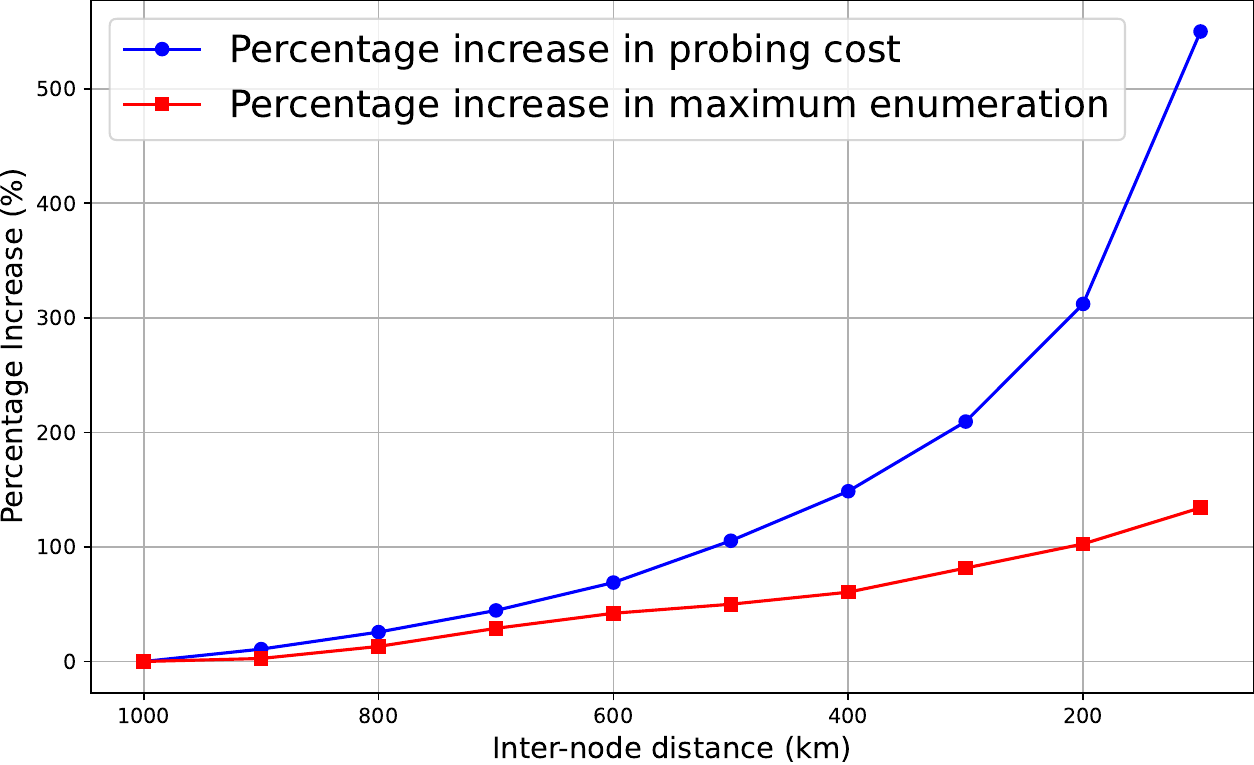}
  \caption{
  Percentage increase in probing cost and enumeration capability when decreasing inter-node distance.
}
  \label{fig:enumeration-ripe}
\end{figure}

However, as discussed in \S\ref{sec:enumeration} we do find that RIPE Atlas achieves better enumeration.
Yet, this comes at a significant price in terms of probing cost.
This can be seen in Figure~\ref{fig:enumeration-ripe},
where we plot the number of PoPs detected for a Cloudflare prefix, with presence in 300+ cities, and the probing cost when decreasing the inter-node distance.
Our RIPE Atlas scan was performed using 481 VPs with at least 100km inter-VP distances to maximize geographical coverage.
By increasing the inter-VP distance up to 1,000km we reduce the number of VPs. 
We find that there is a linear increase in enumeration capabilities, whereas the probing cost shows an exponential increase.

\section{Measuring anycast with CHAOS} \label{app:chaos}
\subsection{Detection}
As mentioned in \S\ref{related_work}, prior work attempted to detect and enumerate anycast using \texttt{CHAOS} queries.
We build on this prior work and leverage \texttt{CHAOS} queries as an opportunity to perform a side-by-side comparison in isolation of the anycast-based approach in \manycast and the built-in GCD approach.
We take our nameserver hitlist as a starting point.
We send \texttt{CHAOS} queries to each target from each of our 32 VPs according to RFC~4892~\cite{chaos_rfc} to the IPv4 and IPv6 addresses on the hitlist and record the responses. 
Then we run a separate synchronised anycast-based measurement with 1 second offsets against the hitlist as well as a separate GCD-based measurement.
In other words: we run both the anycast-based and GCD measurement toward all addresses using the \manycast deployment.
Of the 161~K nameservers probed, we detect 2,762 to be anycast using the anycast-based approach of which 2,371 are also found using GCD.
Counting the number of unique \texttt{CHAOS} records, we find 2,404 out of the 161~K nameservers return multiple distinct records.
However, as noted by Fan et al.~\cite{chaos_traceroute}, these include load-balanced nameservers at a single location that return multiple records.
Taking the intersection of nameservers with multiple \texttt{CHAOS} records and those detected as anycast using the anycast-based methodology, we find 1,082 prefixes.
Conversely, 1,175 probed nameservers do not support \texttt{CHAOS}-class records and 147 are either falsely classified as anycast or configured with the same record replicated at all locations.

We also observe 414 nameservers that are unresponsive to ICMP and TCP, hence the GCD method cannot detect them.
50 of these prefixes are detected to be anycast using the anycast-based method,
we observe they give strong indication of being TPs as they are captured at > 3 VPs.
These 50 include prefixes from well-known operators, such as eBay and Oracle, highlighting the utility of the added protocol support in \manycast.

\begin{figure}[t]
  \centering
  \includegraphics[width=\columnwidth]{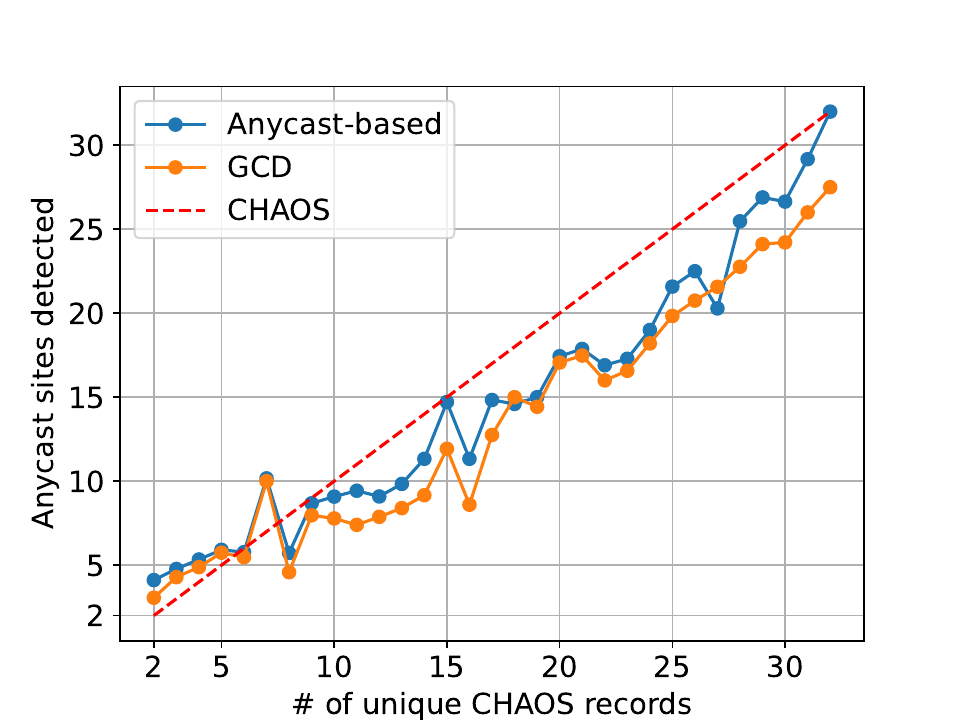}
  \caption{Enumeration counts for the three methodologies: \texttt{CHAOS}, anycast-based, and GCD-based, plotted against the number of unique \texttt{CHAOS} record founds, for IPv4 name servers.
}
  \label{fig:chaos}
\end{figure}
 
 \subsection{Enumeration}
 As mentioned, the \manycast tool has built-in support for both the anycast-based approach and the GCD approach.
 Using the \manycast deployment, with 32 VPs, we measure the site enumeration capabilities of both approaches when using the same VPs.
 We target nameservers that support \texttt{CHAOS} records, and measure the number of distinct \texttt{CHAOS} responsive values observed
under the assumption that different values indicate different sites,
Figure~\ref{fig:chaos} plots the name server IPv4 prefixes by number of unique \texttt{CHAOS} records observed (x-axis) against the site counts found with the different methodologies (y-axis).
We observe two things.
First, for low numbers of distinct \texttt{CHAOS} records, both the anycast- and GCD-based approaches estimate a slightly higher number of sites.
This is unsurprising, as previous work has shown that \texttt{CHAOS} records are often used to differentiate between co-located servers~\cite{chaos_traceroute} (\eg, we observe values such as `auth1' and `auth2' suggesting this behavior)
where they combined it with traceroute to avoid overestimation.
Second, we observe that the anycast-based approach consistently approximates the assumed \texttt{CHAOS} ``ground truth'' more closely suggesting the anycast-based approach may be slightly better at estimating a lower bound for the number of anycast sites, especially for cases where the anycast-based approach receives responses at many vantage points.
Despite \texttt{CHAOS} records being a poor indicator of anycast, we intend on including it in our daily scanning as it provides insightful information for nameservers that implement this.
\section{External dataset}\label{app:bgptools}

\begin{table}
\resizebox{\columnwidth}{!}{
\begin{tabular}{|c|r|r|r|r|}
    \hline
    \textbf{Prefix size} & \textbf{Occurrence} & \textbf{Anycast} & \textbf{Unicast} & \textbf{Unresponsive} \\
    \hline
    \hline
    \textbf{/11} & 1 & 1,026 & 1,845 & 5,321 \\
    \hline
    \textbf{/13} & 1 & 256 & 0 & 1,792 \\
    \hline
    \textbf{/14} & 7 & 841 & 3,844 & 2,483 \\
    \hline
    \textbf{/15} & 2 & 84 & 372 & 568 \\
    \hline
    \textbf{/16} & 16 & 976 & 1,262 & 1,858 \\
    \hline
    \textbf{/17} & 5 & 273 & 211 & 156 \\
    \hline
    \textbf{/18} & 4 & 135 & 54 & 67 \\
    \hline
    \textbf{/19} & 9 & 70 & 87 & 131 \\
    \hline
    \textbf{/20} & 221 & 3,378 & 33 & 125 \\
    \hline
    \textbf{/21} & 16 & 65 & 32 & 31 \\
    \hline
    \textbf{/22} & 51 & 175 & 8 & 21 \\
    \hline
    \textbf{/23} & 134 & 213 & 21 & 34 \\
    \hline
    \textbf{/24} & 2,580 & 2,247 & 269 & 64 \\
    \hline
    \hline
    \textbf{Total} & 3,047 & 9,739 & 8,038 & 12,651 \\
    \hline
\end{tabular}
}
\vspace{0.1em}
\caption{
	BGP prefixes classified as anycast by \textit{BGPTools} grouped by size and its occurrence. We count the number of anycast, unicast, and unresponsive /24s according to GCD.
}
\vspace{-2em}
\label{table:bgptools}
\end{table}

As mentioned in \S\ref{subsec:external}, \textit{BGPTools} uses the anycast-based approach to detect anycast.
When an address is classified as anycast they assume the entire BGP prefix announcing this address is anycast.
In this section we show this assumption does not hold by comparing both daily censuses using data from the 20th of Dec.\ '24.

In Table~\ref{table:bgptools} we show the BGP prefixes detected as anycast in the \textit{BGPTools} census grouped by size.
We see prefixes range from /11 (that contains 8,192 /24s) to /24 in size.
Next, we list the occurrence of each prefix size, we observe the most occurring prefix size is /24 followed with /20.
For these prefix sizes we list the number of anycast, unicast, and unresponsive /24s found in our census using GCD.

First, for the 2,580 \textit{BGPTools} /24-prefixes we find 2,247 are GCD-confirmed, 269 not GCD-confirmed, and 64 are unresponsive.
We suspect the 269 not GCD-confirmed prefixes are largely FPs of the anycast-based approach.
Looking at the total, we observe \textit{BGPTools} classifies 3,047 BGP prefixes as anycast that include 9,739 GCD-confirmed anycast /24s; 8,038 unicast /24s; and 12,651 unresponsive /24s.
For the single /13 prefix we count 256 anycast /24s. Investigating this prefix we observe it contains a /16 that is entirely anycast, whereas the remainder of the /13 is ICMP unresponsive.
Investigating the 7 /14s, that have the most unicast /24s, we find 6 are prefixes announced by Google Cloud.
Using the aforementioned Google \textit{ipranges} dataset, we confirm these unicast /24s are not listed as being announced globally.
These results reaffirm that GCD confirmation is necessary to filter out FPs of the anycast-based approach and that their assumption to classify entire BGP prefixes as anycast is wrong.

\begin{figure}[b]
  \centering
  \includegraphics[width=\columnwidth]{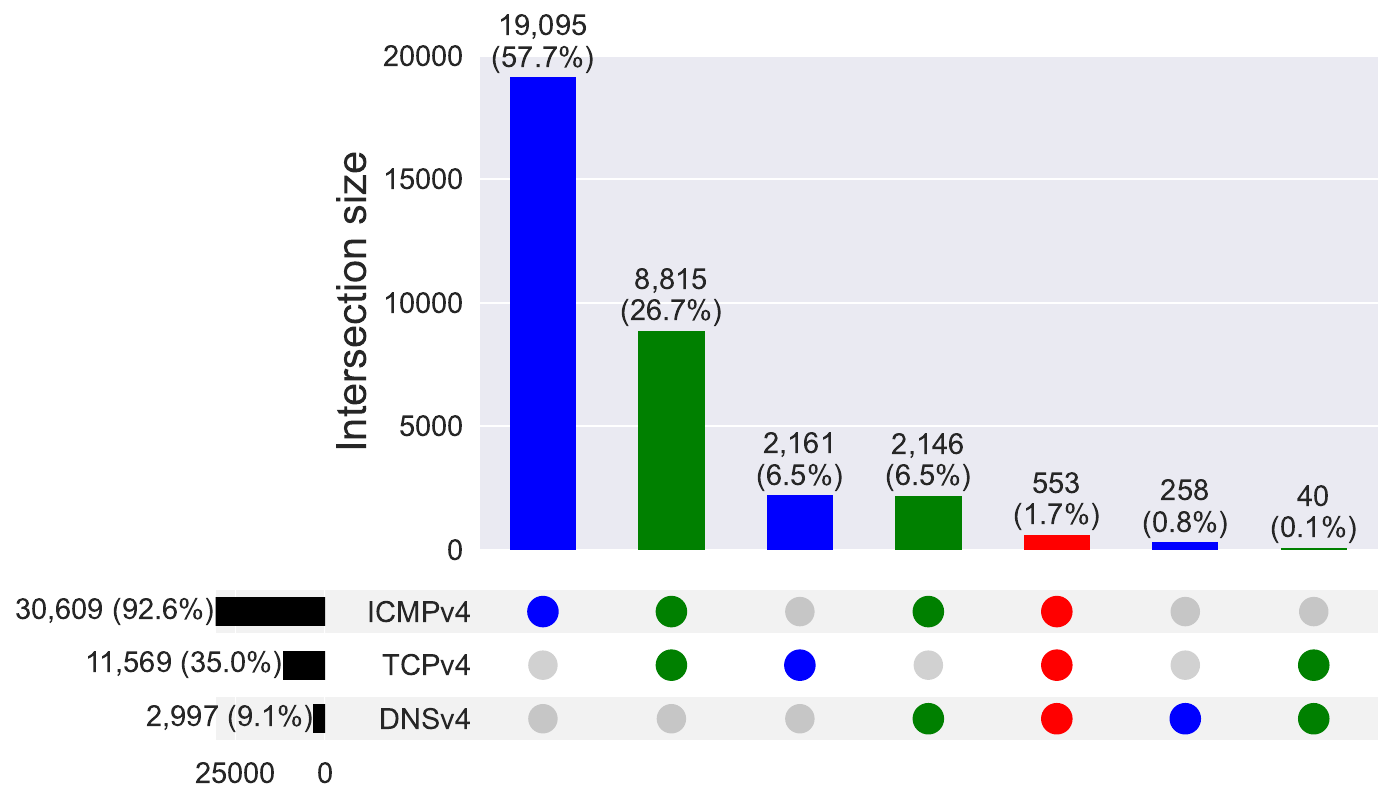}
  \caption{\manycast~ detection of anycast candidates for ICMPv4, TCPv4, and DNSv4.}
  \label{fig:upset}
\end{figure}

\begin{figure}[b]
  \centering
  \includegraphics[width=\columnwidth]{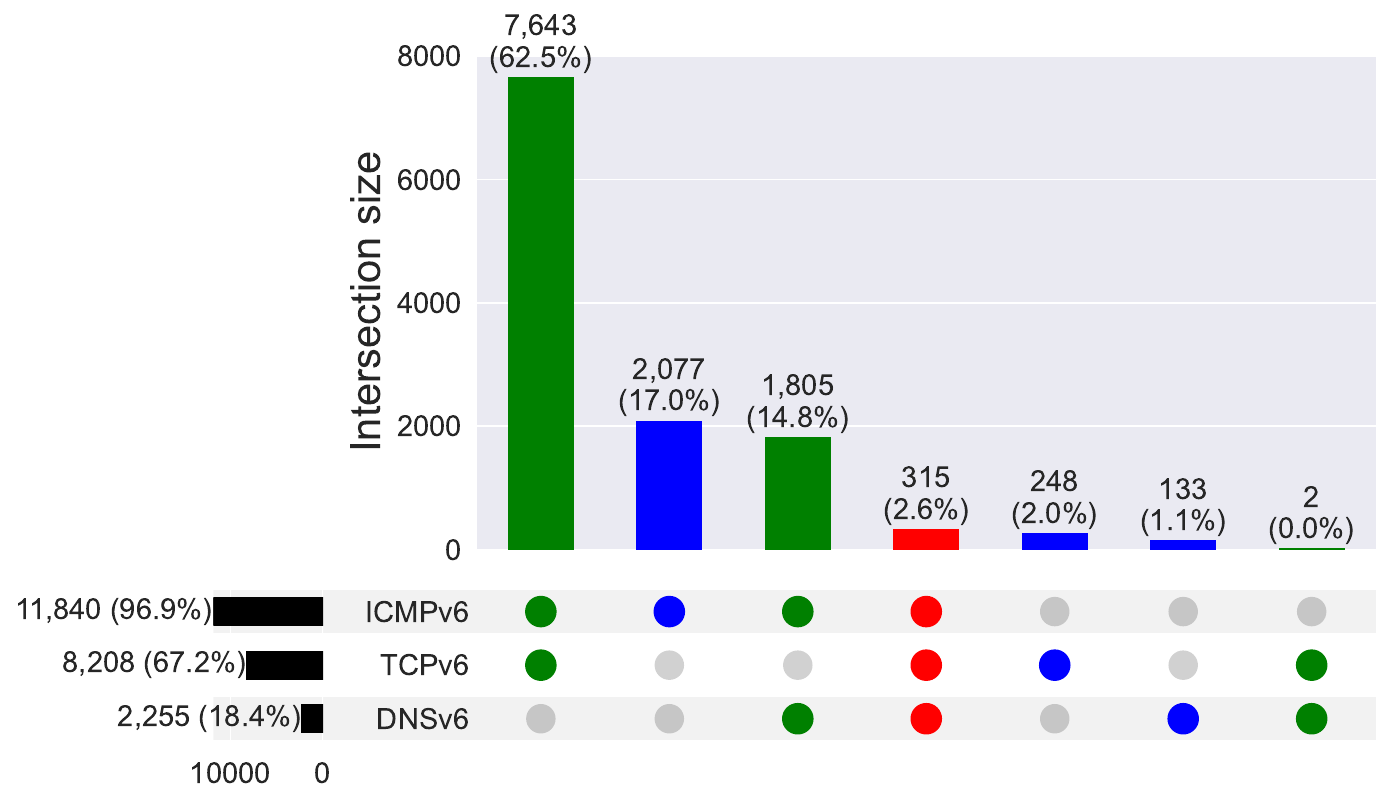}
  \caption{\manycast~ detection of anycast candidates for ICMPv6, TCPv6, and DNSv6.} 
  \label{fig:upsetv6}
\end{figure}

\section{Protocol coverage}\label{app:protocols}
To extend coverage compared to iGreedy and \manycasttwo we added TCP and DNS over UDP probing.
Figure~\ref{fig:upset} shows a breakdown of which prefixes can be detected using our anycast-based approach with different protocols using a so-called \textit{upset} plot.
First, the bottom left shows the totals for each protocol, \ie, we detect 30,609 anycast prefixes with ICMP,
11,569 with TCP, and 2,997 with DNS.
Next, below the histogram we indicate the intersection shown by each bar. We color this by three categories: \textbf{blue} for results of a single protocol, \textbf{green} for groupings of two protocols, and \textbf{red} for the intersection of all three protocols.
The bars show the totals and percentage that each intersection gives,
\eg, the first bar shows that 19,095 prefixes (57.7\% of the total) were discovered to be anycast using \textbf{only} ICMP (\ie, TCP and DNS did not detect these prefixes as anycast).
Next, we have the prefixes detected as anycast by both ICMP and TCP, totaling to 8,815, etc.
We also include the IPv6 Upset plot in Figure~\ref{fig:upsetv6}.

\section{Disclaimer}
Any views and opinions expressed in this work are those of the authors and do not necessarily reflect those of the European Union, the European Research Council, NSF, or the Dutch research funding agency NWO. Neither the European Union, nor NSF, nor NWO can be held responsible for them.

\end{document}